\documentclass[sigconf]{acmart}
\renewcommand\footnotetextcopyrightpermission[1]{}
\usepackage{booktabs} % For formal tables

\usepackage{xcolor}

\usepackage{algpseudocode}
\usepackage{algorithm}
\usepackage{amsmath}
\usepackage{amssymb}
\usepackage{url}

\usepackage{graphicx}
\graphicspath{ {images/} }

\usepackage{caption}
\usepackage{multicol}
\usepackage[T1]{fontenc}
\usepackage{diagbox}

\usepackage{subcaption}

\usepackage[acronym]{glossaries}

\usepackage[inline]{enumitem}

\newacronym{gao}{Gao's}{The Gao's algorithm}
\newacronym{mvp}{MVP}{The MVP algorithm}
\newacronym{ucla}{UCLA}{The UCL algorithm}
\newacronym{asrank}{ASRANK}{The ASRANK algorithm}
\newacronym{pari}{PARI}{The PARI algorithm}
\newacronym{mlr}{MLR}{Multinomial Logistic Regression}
\newacronym{principle}{\textcircled{1}}{Step~\textcircled{1}}
\newacronym{probabilistic}{\textcircled{2}}{Step~\textcircled{2}}

\makeatletter

\newcommand{\Rmnum}[1]{\expandafter\@slowromancap\romannumeral #1@}
\makeatother

\begin{document}
	\title{\acrshort{pari}: A \underline{P}robabilistic Approach to \underline{A}S \underline{R}elationships \underline{I}nference}
	
	\author{Guoyao~Feng}
	\affiliation{%
		\institution{Carnegie Mellon University}
	}
	\email{gfeng@cs.cmu.edu}

	\author{Srinivasan~Seshan}
	\affiliation{%
		\institution{Carnegie Mellon University}
	}
	\email{srini@cs.cmu.edu}

	\author{Peter~Steenkiste}
	\affiliation{%
		\institution{Carnegie Mellon University}
	}
	\email{prs@cs.cmu.edu}
	
	\begin{abstract}
		Over the last two decades, several algorithms have been proposed to infer the type of relationship between Autonomous Systems (ASes). While the recent works have achieved increasingly higher accuracy, there has not been a systematic study on the uncertainty of AS relationship inference. In this paper, we analyze the factors contributing to this uncertainty and introduce a new paradigm to explicitly model the uncertainty and reflect it in the inference result. We also present \acrshort{pari}, an exemplary algorithm implementing this paradigm, that leverages a novel technique to capture the interdependence of relationship inference across AS links.
	\end{abstract}
	
	\settopmatter{printfolios=true}
	\maketitle
	
	\section{Introduction}
\label{sec:intro}
The Internet inter-domain routing topology consists of thousands of independent interconnected autonomous systems (AS). To route traffic in the Internet, these ASes connect to each other through the Border Gateway Protocol (BGP) and establish routing relationships, which are based on their private business agreements. 
Knowing these business relationships is essential to understanding how packets are routed on the network today and, therefore, critical to evaluating new protocol designs or network configurations. Unfortunately, these business agreements are typically kept private, which has motivated numerous past efforts~\cite{gao2001inferring,xia2004evaluation,subramanian2002characterizing,di2003computing,dimitropoulos2007relationships,oliveira2010completeness,luckie2013relationships} to infer the relationships based on observations of network routing. 
Past efforts typically classify the relationship between two ASes into one of the three major types: customer-to-provider (c2p)~\footnote{Inferring an AS link $u-v$ as c2p is equivalent to inferring the link $v-u$ as p2c.}, peer-to-peer (p2p), and sibling-to-sibling (s2s). 
While the accuracy of these inference algorithms has improved over time, we argue that inference algorithms inherently incorporate some form of uncertainty and error. In this paper, we argue that, unlike past work, explicitly incorporating uncertainty into reasoning about the AS relationships provides valuable benefits. 

One source of the uncertainty in reasoning about AS relationships is that the fundamental guiding assumption adopted by the prior algorithms---the valley-free principle~\cite{gao2001inferring}---is by itself inadequate to deterministically infer the relationship for every link in the AS-level topology~\cite{dimitropoulos2007relationships}. For instance, the principle can never infer a link as p2p and eliminate the possibility of it being p2c. As a result, past work must resort to additional assumptions and heuristics to classify links. Past efforts do consider the reliability of these heuristics ``implicitly'' by simply rank ordering them. This concept of relative uncertainty is never conveyed in the final output or used to holistically reason about the decision. In addition, the inference results depend on the coverage of AS paths received by the route collectors. For example, some algorithms~\cite{luckie2013relationships} rely on ASes' transit/node degrees as a hint to determine the provider in a p2c relationship. However, newly deployed collectors may uncover more AS paths, causing ASes' transit/node degrees and inferred relationships to change. 

Motivated by the challenges above, we propose a novel framework that allows the algorithmic implementation to explicitly express and model uncertainty. Our framework features two key steps: \textit{principle-based} and \textit{probabilistic} inference. The principle-based inference step applies only dependable rules (principles) to make deterministic relationship inferences and ensures the inference results remain stable under varying coverage of AS paths. It takes a conservative strategy to trading off coverage in the number of links inferred for high accuracy. Then, for each undecided link, the probabilistic inference step outputs a vector of scores rather than a single relationship. This score represents our confidence in inferring the link as one of the possible relationships. Thus, our paradigm embraces uncertainty and makes it explicit when there is no strong evidence in favor of a particular type of relationship. 
 
We have designed a concrete algorithm, ~\acrshort{pari}, that illustrates the value of this framework. ~\acrshort{pari} uses two principles for the principle-based step. The first principle leverages the observation that there exists a clique of well-known Tier-1 ASes interconnected by p2p links~\cite{oliveira2010completeness} and the second one is the valley-free principle. For probabilistic inference, we develop a novel technique to capture the interdependence between undecided links. For each link, it computes \textit{share}, a metric that measures the number of valley-free inferences over the entire topology, under the condition that the link is inferred as one of the possible relationships. Then~\acrshort{pari} uses the links' shares as the features of a machine learning algorithm (logistic regression) to compute probability estimates as the score output of inference. It leverages a ground-truth data set to learn the model parameters.

Our evaluation results show that~\acrshort{pari}'s principle-based inference achieves higher accuracy and is more stable than prior algorithms. Moreover,~\acrshort{pari}'s probabilistic inference outperforms a na\"ive approach to expressing uncertainty. We also evaluate the efficacy of the share metric by comparing it against other metrics such as transit and node degrees as features for learning. We show that the share metric is not only effective by itself but also complementary to the transit and node degrees.

The contributions of our work are summarized as follows:
\begin{enumerate}
	\item We identify 4 sources of uncertainty in AS relationship and analyze their impact on the inference results.
	\item We propose a novel framework to model inference uncertainty explicitly and reflect it in the results.
	\item We propose PARI, an exemplary algorithm that implements this framework and leverages a novel technique to capture the interdependence between undecided links. Our evaluation indicates it achieves higher accuracy and stability compared to prior algorithms for deterministic inference and outperforms a na\"ive strategy to express uncertainty.
\end{enumerate}

The paper is organized as follows. \S\ref{sec:background} describes the related works on AS relationship inference. In \S\ref{sec:uncertainty} we provide insights into the causes of uncertainty in relationship inference. \S\ref{sec:rationale} describes the new paradigm to explicitly reflect uncertainty in the inference output, followed by \S\ref{sec:pari} that presents the \acrshort{pari} algorithm as an exemplary implementation of the paradigm. We evaluate the performance of \acrshort{pari} in \S\ref{sec:evaluation}. \S\ref{sec:conclusion} concludes our work.
	\section{Related Work and Background}
\label{sec:background}

\textbf{Inference Algorithms.} The topic of AS relationship inference was started by Gao's pioneering work~\cite{gao2001inferring} that classified an AS link into one of the three relationship types: p2c, p2p, and s2s. In a c2p relationship, the customer AS pays the provider AS for the traffic transmitted between them to obtain global reachability. In a p2p relationship, the two parties transmit traffic between their own networks and their customers networks with no payment involved. In the case of s2s relationship, the two ASes are owned by the same organization, and they transmit traffic between their providers/peers/siblings for free.

New inference algorithms~\cite{subramanian2002characterizing, di2003computing, dimitropoulos2007relationships, oliveira2010completeness, luckie2013relationships} have been proposed in the follow-up works to achieve increasingly high accuracy. But \textbf{uncertainty} arises in relationship inference when they face problems such as the incomplete coverage of AS-level topology and the use of potentially unreliable heuristics. Unfortunately, none of the prior algorithms explicitly address it. Our work thus seeks to identify these contributing factors and develop a new paradigm that explicitly models the uncertainty and reflects it in the inference result.

We give a brief description of prior algorithms to provide the necessary context for the following sections. \acrshort{gao} algorithm assumes the \textit{valley-free} property in AS paths to guide its inference. The valley-free property requires that an AS path starts with an uphill segment consisting of zero or more c2p or s2s links, followed by zero or one p2p links, and a downhill segment formed by zero or more p2c or s2s links. It also assumes a provider is usually larger than its customers and peers are of similar size in terms of node degree. 

\cite{subramanian2002characterizing} formally defines the Type of Relationship (ToR) problem and presents a heuristic-based algorithm (MVP) that exploits partial views of the AS graph available from different vantage points. The ToR problem models relationship inference as an edge labeling problem on a graph derived from the AS paths, and asks for a labeling that maximizes the number of valley-free AS paths.
\cite{di2003computing} proves the speculation~\cite{subramanian2002characterizing} that the ToR problem is indeed NP-complete in general. Their approach (DPP) involves mapping a simplified variant of the ToR problem to a 2SAT~\cite{garey1979computers} problem and is shown to be tractable if all AS paths follow the valley-free constraint.

The algorithm proposed in \cite{dimitropoulos2007relationships} starts by inferring s2s relationships based on information from the IRR databases. Then it casts the task of c2p inference into a MAX2SAT problem that attempts to simultaneously maximize the number of valley-free paths and minimize the number of degree inversions (i.e., customer having a higher node degree than the provider). The last step infers p2p links using a heuristic based on~\cite{di2003computing} and~\cite{gao2001inferring}.

The UCLA algorithm in~\cite{oliveira2010completeness} is designed based on the completeness of the AS-level topology observed by the public view. It obtains a clique of all Tier-1 ASes from external sources and infer as p2c the links following these ASes along the AS paths. The remaining links are inferred as p2p.

The \acrshort{asrank} algorithm in~\cite{luckie2013relationships} does not strictly follow the valley-free constraint as the constraint's underlying assumption about routing decision is not always valid~\cite{roughan201110}. Instead, it is replaced by the following assumptions: (1) there exists a peering clique of large transit providers at the top of the hierarchy; (2) a provider will announce customers' routes to its providers; and (3) cycles of p2c links are not allowed. Their ground truth dataset covers 34.6\% of the inferred relationships and shows \acrshort{asrank} correctly infers 99.6\% of known c2p relationships and 98.7\% of known p2p relationships. Our goal is not to dramatically improve upon this measure of accuracy. Instead, it is to quantify the confidence that such systems have in their output decisions.

\textbf{Data.} We downloaded BGP data from three archives: PCH~\cite{pch}, RIS~\cite{ripe}, and Routeviews~\cite{routeviews}. PCH (Packet Clearing House) manages route collectors deployed on more than 100 IXPs around the world. It makes publicly available the daily routing table snapshots from these collectors. RIS (Routing Information Service), developed by RIPE NCC, also collects and stores Internet routing data from several locations around the globe. Routeviews started as a tool to obtain real-time information about the global routing system from the perspectives of several different backbones and locations around the Internet. It maintains a data archive of BGP RIBs and updates. Each collector manages multiple vantage points and receives their BGP data feeds.

The BGP dataset used by our experiments contained the AS paths from the daily routing table snapshots between April 1, 2012 and April 5, 2012. We chose this time range to match the timestamp of the ground truth dataset for evaluation, as we describe below. We combine AS paths from all collectors into the single \textit{aggregate} path set, $Q$, which is then used as the input to the inference algorithms. We use $I$ to denote the set of all collectors. $I$ captures 79 route collectors with at least one valid AS route, over 128k AS links and more than 41k ASes.

We use the \textit{partial} ground truth dataset released by~\cite{luckie2013relationships}. It is derived from community attributes~\cite{chandra1996bgp} and RPSL~\cite{alaettinoglu1999routing} from April 2012. It covers 16,248 p2p links and 31,886 p2c/c2p links.

We use the datasets above for the rest of the paper.
	\section{Uncertainty in AS Relationship Inference}
\label{sec:uncertainty}
In this section, we present 4 factors (\S\ref{sec:stability}-\ref{sec:flaky}) that make it impossible to deterministically infer a single relationship for certain links and describe how these factors can influence an algorithm's output. We also provide some insights into possible strategies to mitigate their impact, paving the way for an uncertainty-aware solution. At the end we discuss the potential benefits of exposing inference uncertainty.

\subsection{Dependence on Route Collectors' Coverage}
\label{sec:stability}

The ability to deploy route collectors can place limits on our visibility of the AS-level Internet topology. 
The incompleteness of the observable AS-level Internet topology has been studied by past research efforts. 
For example, \cite{oliveira2008search, oliveira2010completeness} showed that single-period snapshots of public BGP datasets only reveal a small portion of the links connecting Tier-2 ASes; combining long period snapshots and historical data of BGP updates leads to significant improvement. Moreover, the public BGP data misses many peer links at Tier-2 and below, as a result of the valley-free property and the poor coverage of monitors at stub ASes. In this section, we analyze how these coverage limitations influence the results of inference algorithms. 
Specifically, \S\ref{sec:case-study} illustrates the effect through concrete examples drawn from prior algorithms: \acrshort{gao}, \acrshort{asrank}, and \acrshort{mvp}. \S\ref{sec:stability-eval} presents an empirical study using public BGP datasets.

Let us define the key terms used in the rest of the paper. The \textit{vantage points} of a collector are identified by the set of first-hop ASes along the AS paths it receives. In Figure~\ref{fig:gao-example}, vantage point $A$ is feeding BGP routing data (i.e., path $A-B-C-D$) to its route collector $RC$.

We quantify a route collector's \textit{coverage} as the number of links and vantage points it is able to observe\footnote{We choose not to measure by the number of unique paths because paths traversing the top of the hierarchy tend to share the same segments and contribute little new information in terms of coverage. For example, the same path announcement received by two vantage points constitute two unique paths even though the only difference is the ASN prepended by the vantage points.}. We refer to the undirected graph formed by the set of AS paths from a set of vantage points as the \textit{Original} AS topology whereas the one with additional vantage points as the \textit{Augmented} AS topology. Given an AS link shared by both, an inference algorithm is deemed \textit{unstable} if it infers different relationships under the two topologies, which implies inference uncertainty due to route collectors' limited coverage.

\subsubsection{Instability Causes}
\label{sec:case-study}
Algorithms can be unstable when paths from new vantage points becomes available as input, if they infer relationships based on measures that involve aggregates such as the total number of adjacent ASes. 

We use \acrshort{gao} algorithm as an example to illustrate the problem. \acrshort{gao} algorithm starts by computing the node degree of each AS in the topology. Next, for each AS path, one of the ASes with the highest node degree is selected as the peak of the valley-free path. It then increments the c2p score of links preceding the peak and the p2c score of links following the peak. The third step infers links as p2c/c2p if their scores exceed a user-defined threshold. Finally, the algorithm revisits each AS path, filters out links that are impossible to have peering relationships, and assigns p2p relationship to links whose end points have similar node degrees.

\begin{figure}[htb!]
	\centering
	\includegraphics[width=0.9\linewidth]{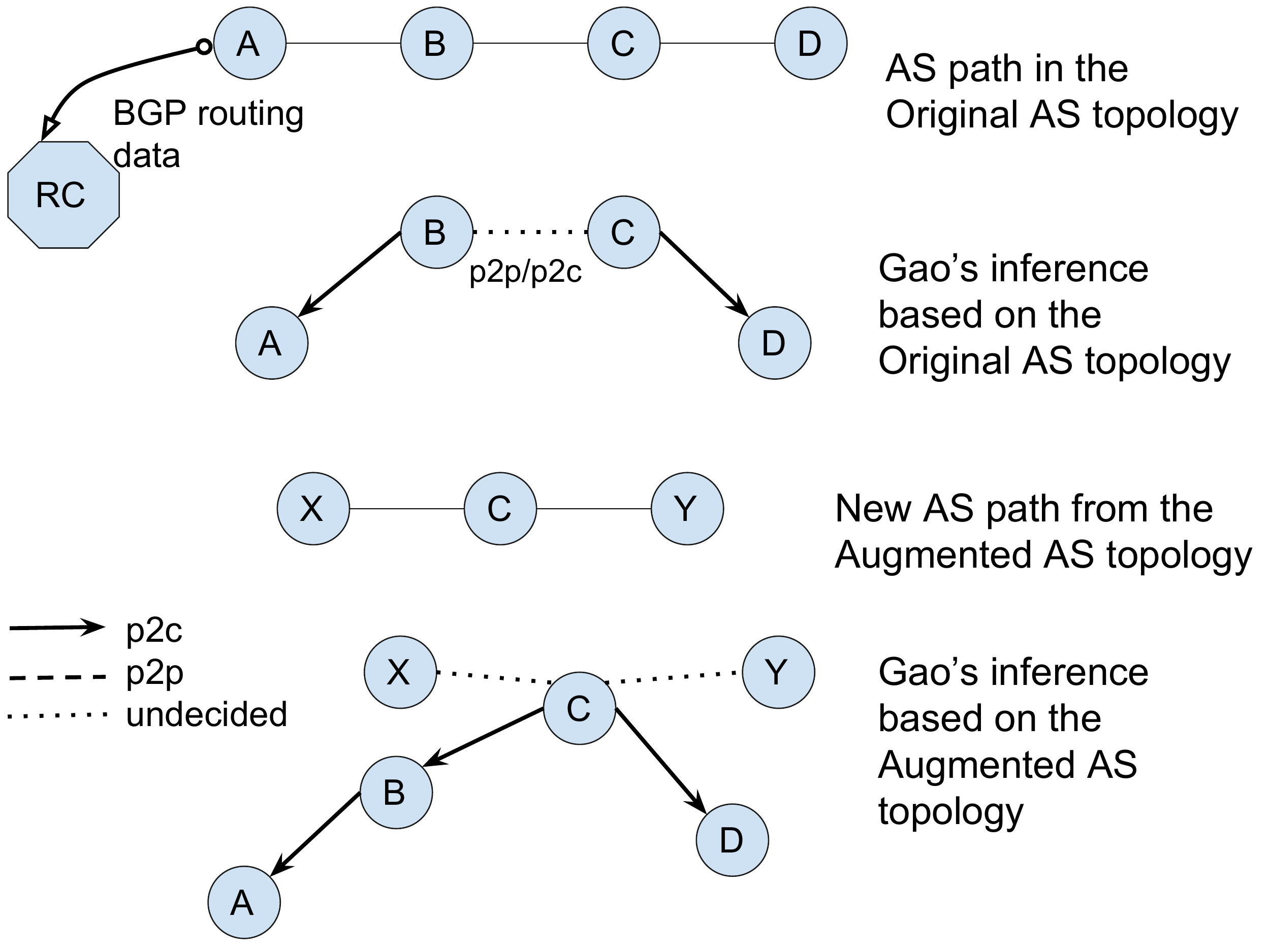}
	\caption{An example illustrating the impact of incomplete coverage on \acrshort{gao} algorithm's stability. In the Original topology, either B or C is the top provider. C becomes the top provider after a new path is introduced by the Augmented topology}
	\label{fig:gao-example}
\end{figure}

To see how incomplete coverage affects the output of \acrshort{gao} algorithm, consider the AS path $A-B-C-D$ in the Original topology in Figure~\ref{fig:gao-example}. Both $B$ and $C$ have a node degree of 2.~\acrshort{gao} algorithm chooses one of the ASes with the highest node degree as the peak of the valley-free path; ASes on both sides of the peak are classified as either peers or customers. Without loss of generality we assume it selects $B$ as the peak of the AS path and $B-C$ is classified as either p2c or p2p. Now consider a new path in the Augmented topology, $X-C-Y$, which leads to $C$'s node degree to increase from 2 to 4. $C$ is now chosen as the peak in the first AS path and the link $B-C$ changes from p2c/p2p to c2p.

Note that the addition of vantage points also impacts other algorithms in similar ways. For example, transit degree, used by the~\acrshort{asrank} algorithm, and vantage-point rank, used by the \acrshort{mvp} algorithm, are both affected by the number of vantage points. 

\subsubsection{Empirical Study of Stability}
\label{sec:stability-eval}
We present an empirical study to evaluate the stability of \acrshort{gao}, \acrshort{mvp}, \acrshort{ucla} and \acrshort{asrank} algorithms.

\begin{figure}
	\begin{subfigure}[b]{.24\textwidth}
		\centering\captionsetup{width=.9\linewidth}
		\includegraphics[width=\textwidth]{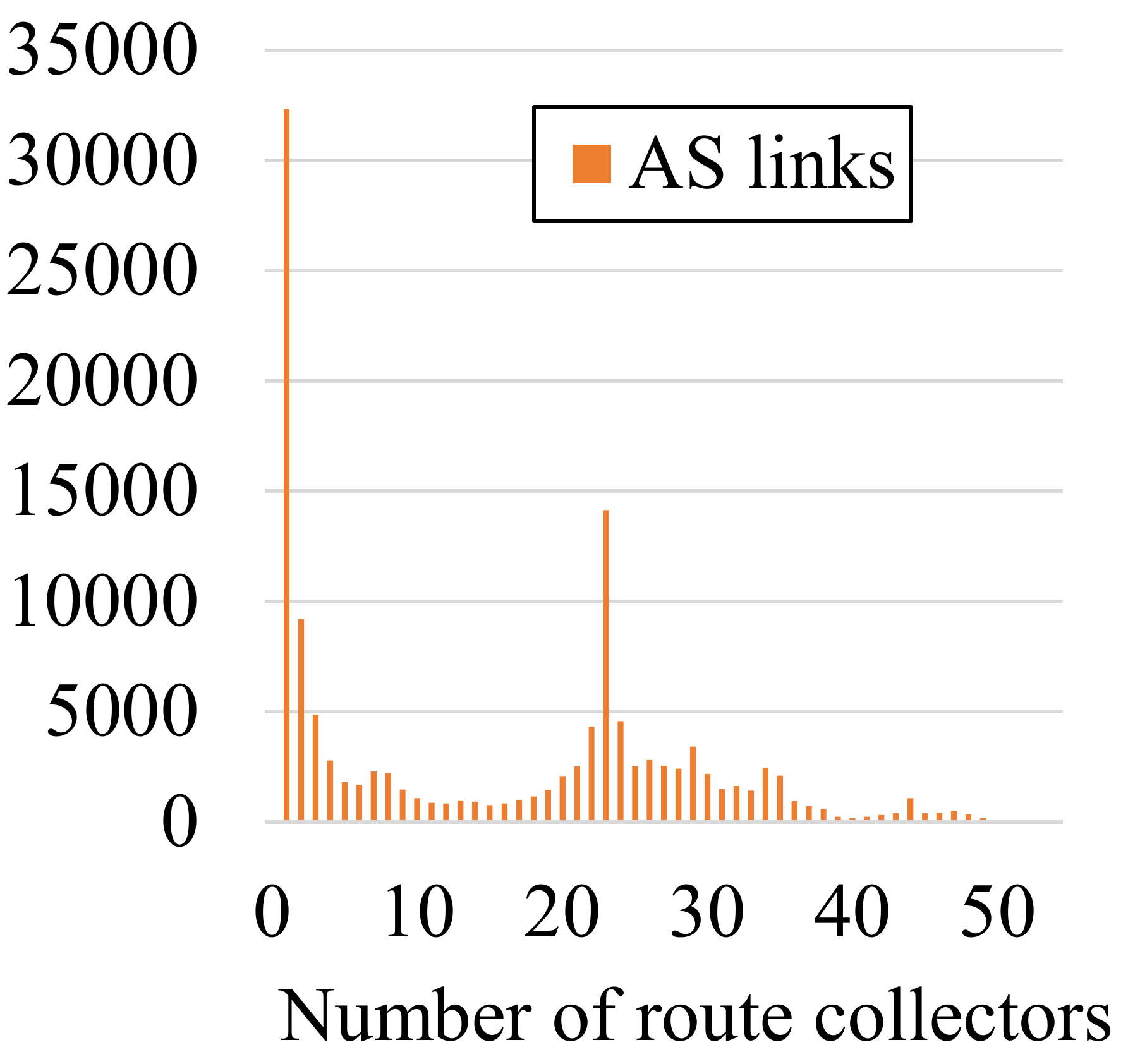}
		\subcaption{Coverage of AS links. 32,322 of them are observable by only one route collector.}
		\label{fig:link-visibility}
	\end{subfigure}%
	\begin{subfigure}[b]{.24\textwidth}
		\centering\captionsetup{width=.9\linewidth}
		\includegraphics[width=\textwidth]{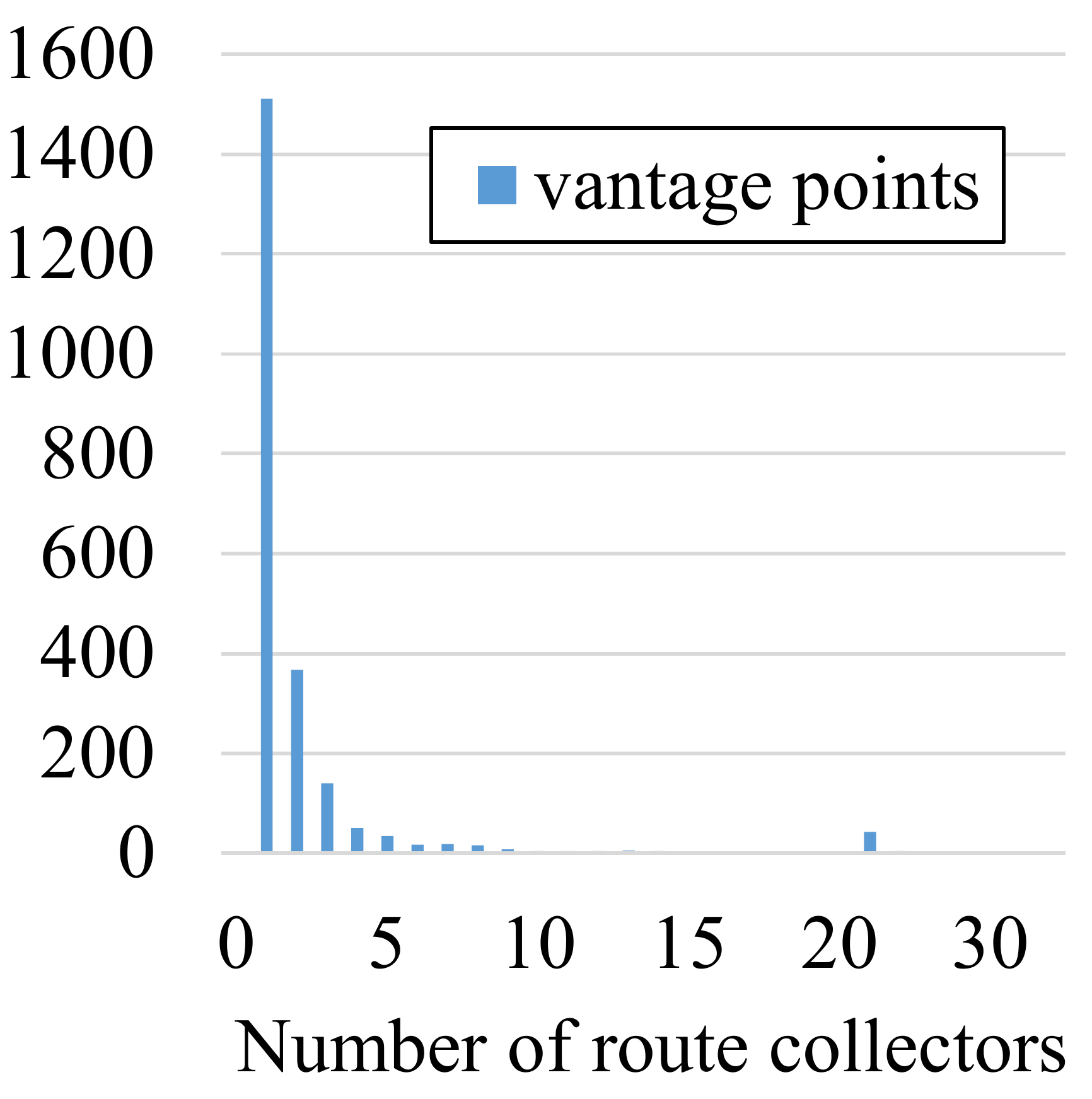}
		\subcaption{Coverage of vantage points. 1,511 of them are observable by only one route collector.}
		\label{fig:source-as-visibility}
	\end{subfigure}
	\caption{Route collectors' coverage of the AS topology.}
	\label{fig:fig}
\end{figure}

\textbf{Route Collector Coverage.}
We start by profiling the coverage of the AS topology by route collectors. The histogram in Figure~\ref{fig:link-visibility} groups the AS links by the number of collectors that are able to observe these links. We note that a significant portion of the links are seen by no more than 2 collectors. They are likely connected to stub ASes near the edge of the AS topology. The spike at 23 can be attributed to the high variance in the number of AS links observed by individual collectors. In particular, 53 collectors observe fewer than 30,000 links while 22 collectors observe over 50,000 links. A similar distribution is found in Figure~\ref{fig:source-as-visibility} which captures the collectors' coverage of vantage points and a minor spike takes place at 21.

The variation in route collectors' coverage has a key implication. While individual collectors might cover some links or vantage points in common, each of them still contributes complementary information to form a more complete view of the AS topology. Moreover, topological information is lost if a route collector is not deployed or becomes unavailable because the majority of links and vantage points are only seen by a single collector.

\textbf{Varying Topology Coverage.}
We evaluate the stability of relationship inference by progressively excluding route collectors from the set of all collectors, $I$. This approach allows us to study the behavior of inference algorithms when existing collectors become unavailable or new collectors are introduced. Suppose $H$ denotes a set of indices of excluded collectors, the remaining collectors is therefore $\overline{H} = I \setminus H$. For example, if $H=\{rrc15\}$, $\overline{H}$ includes all collectors but ``rrc15''. We refer to this step as \textit{shrinking}. We bootstrap $H$ by generating a small random sample from the set of all collectors and gradually grows its by adding more random samples from the remaining collectors. In total, we randomly produce 20 series of $\overline{H}$ for the results below.

A collector never returns to $\overline{H}$ once it is moved to $H$. We measure the stability of inference algorithms as the number of \textit{relationship transitions}, or the number of links whose inference result changes due to shrinking. 

\begin{figure*}[!htb]
	\begin{subfigure}[b]{.33\textwidth}
		\centering\centering\captionsetup{width=.9\linewidth}
		\includegraphics[width=\textwidth]{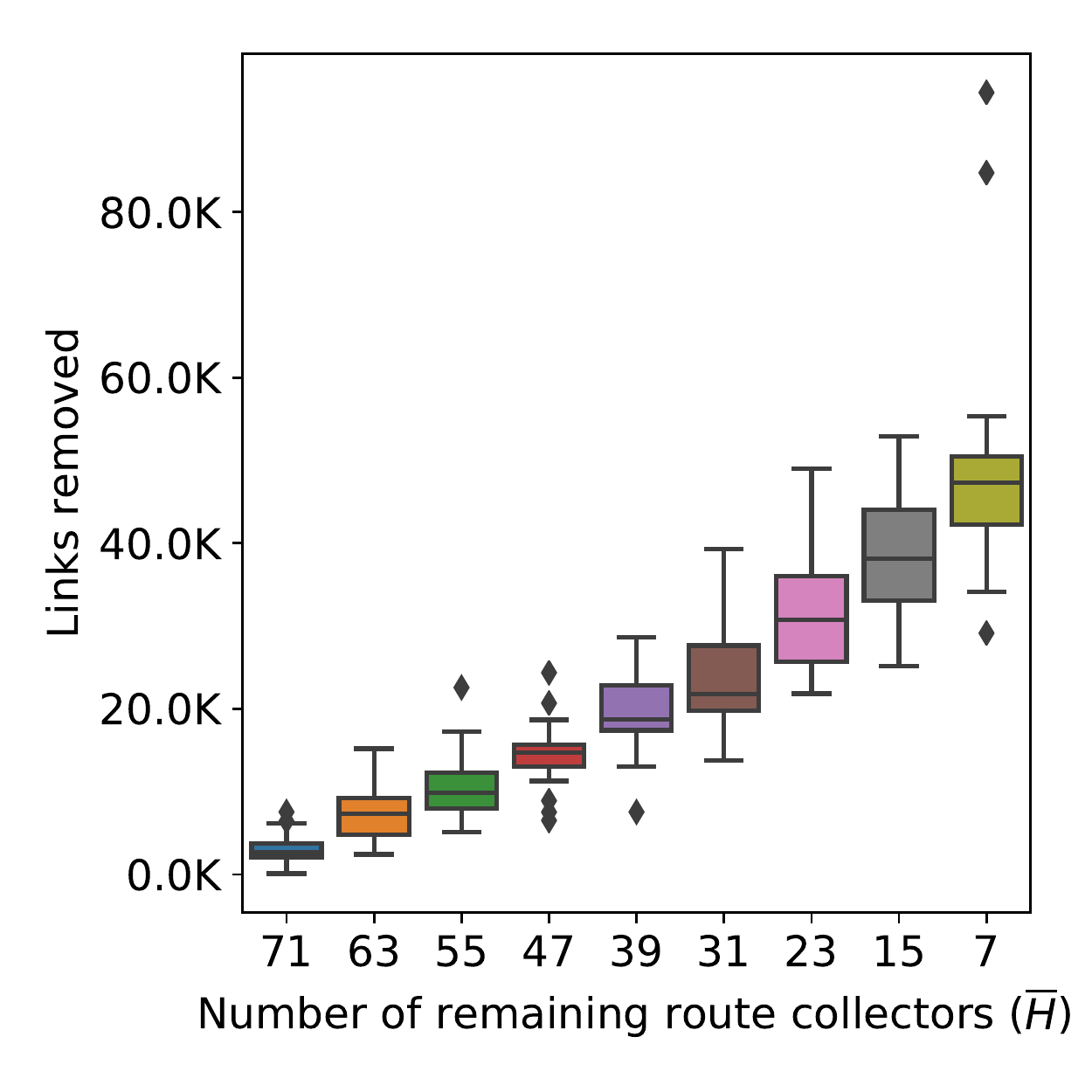}
		\subcaption{Number of links removed as $\overline{H}$ shrinks.}
		\label{fig:links-removed}
	\end{subfigure}%
	\begin{subfigure}[b]{.33\textwidth}
		\centering\centering\captionsetup{width=.9\linewidth}
		\includegraphics[width=\textwidth]{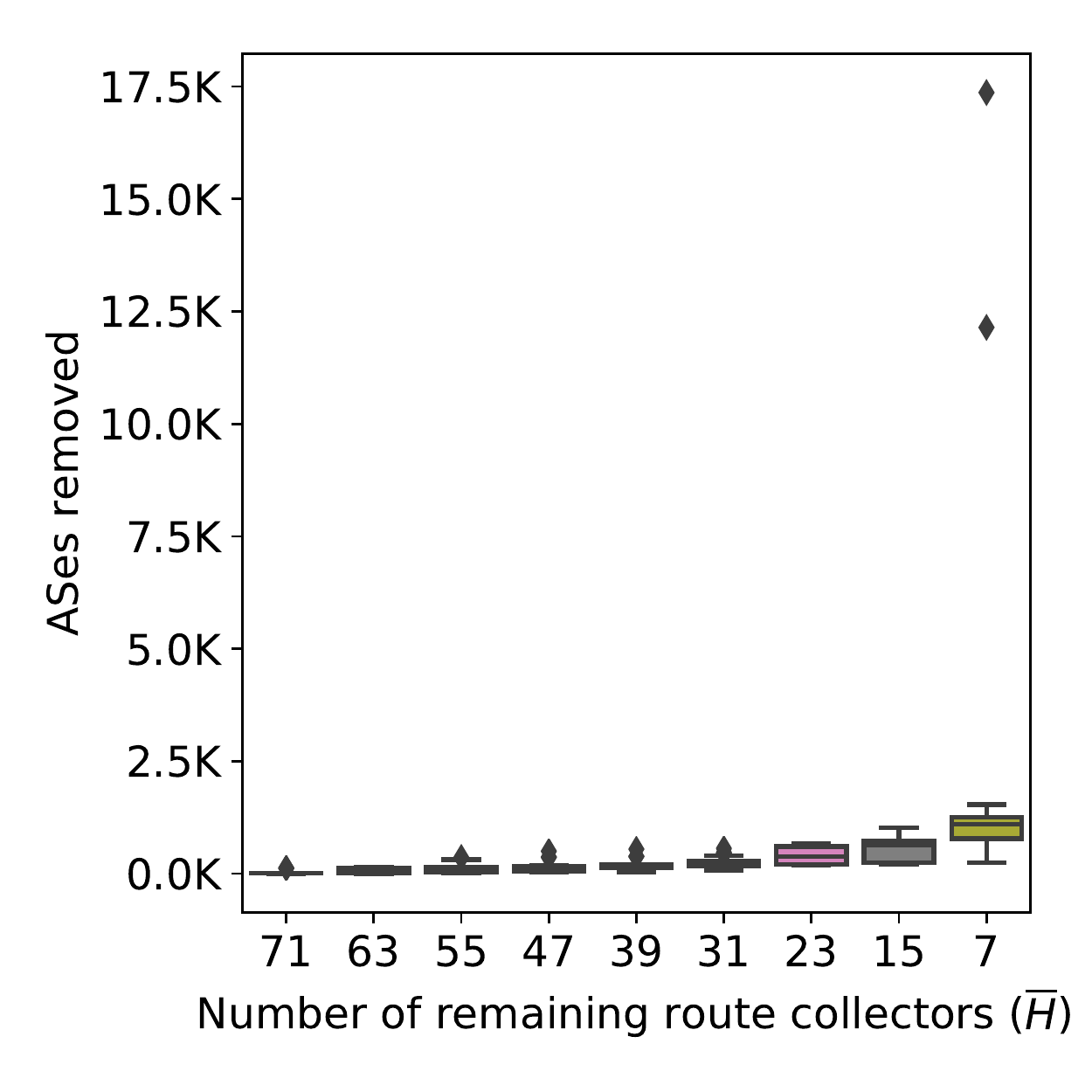}
		\subcaption{Number of ASes removed as $\overline{H}$ shrinks.}
		\label{fig:ASes-removed}
	\end{subfigure}%
	\begin{subfigure}[b]{.33\textwidth}
		\centering\captionsetup{width=.9\linewidth}
		\includegraphics[width=\textwidth]{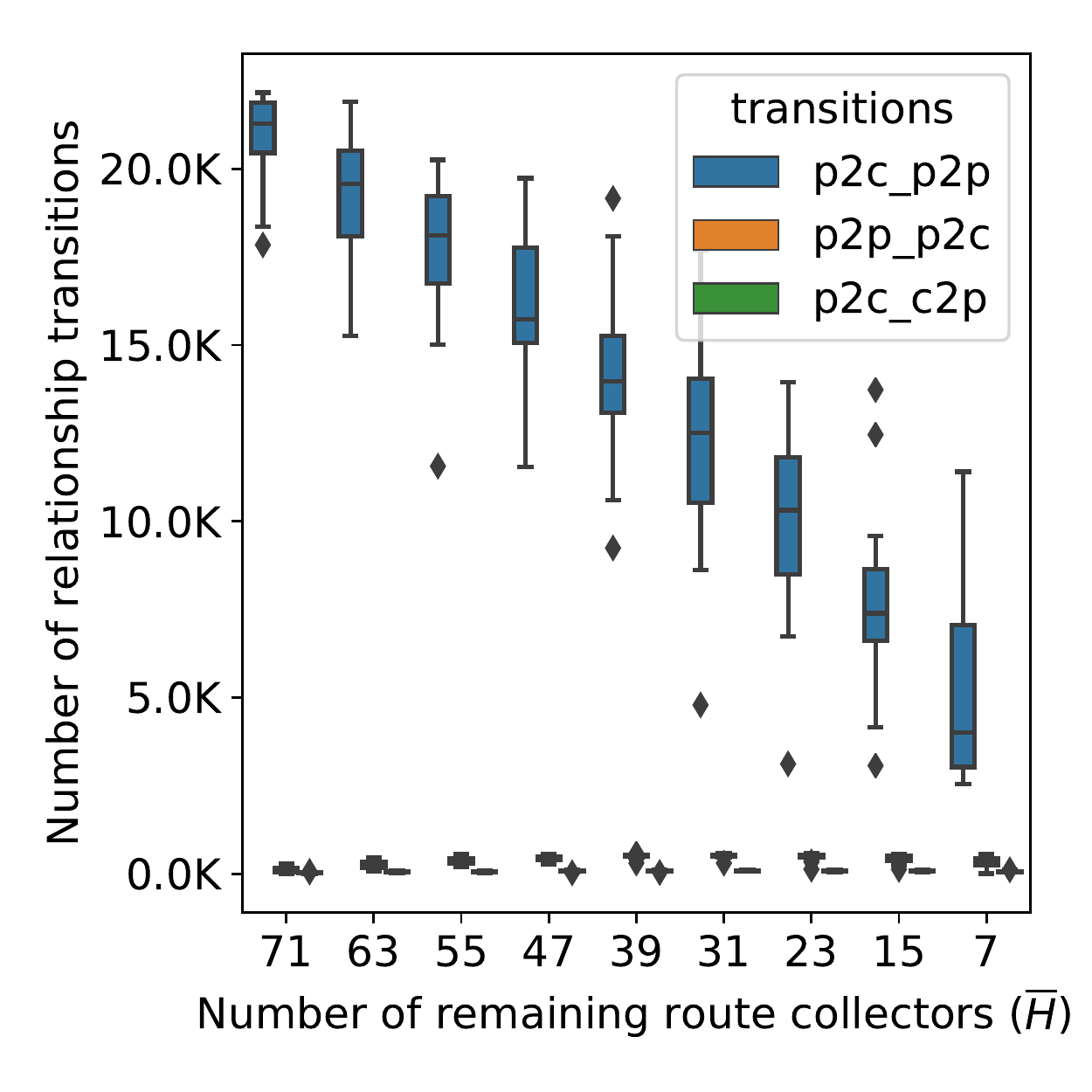}
		\subcaption{Relationship transitions as a function of $\overline{H}$ for \acrshort{gao}.}
		\label{fig:gao-transitions}
	\end{subfigure}
	
	\begin{subfigure}[b]{.33\textwidth}
		\centering\captionsetup{width=.9\linewidth}
		\includegraphics[width=\textwidth]{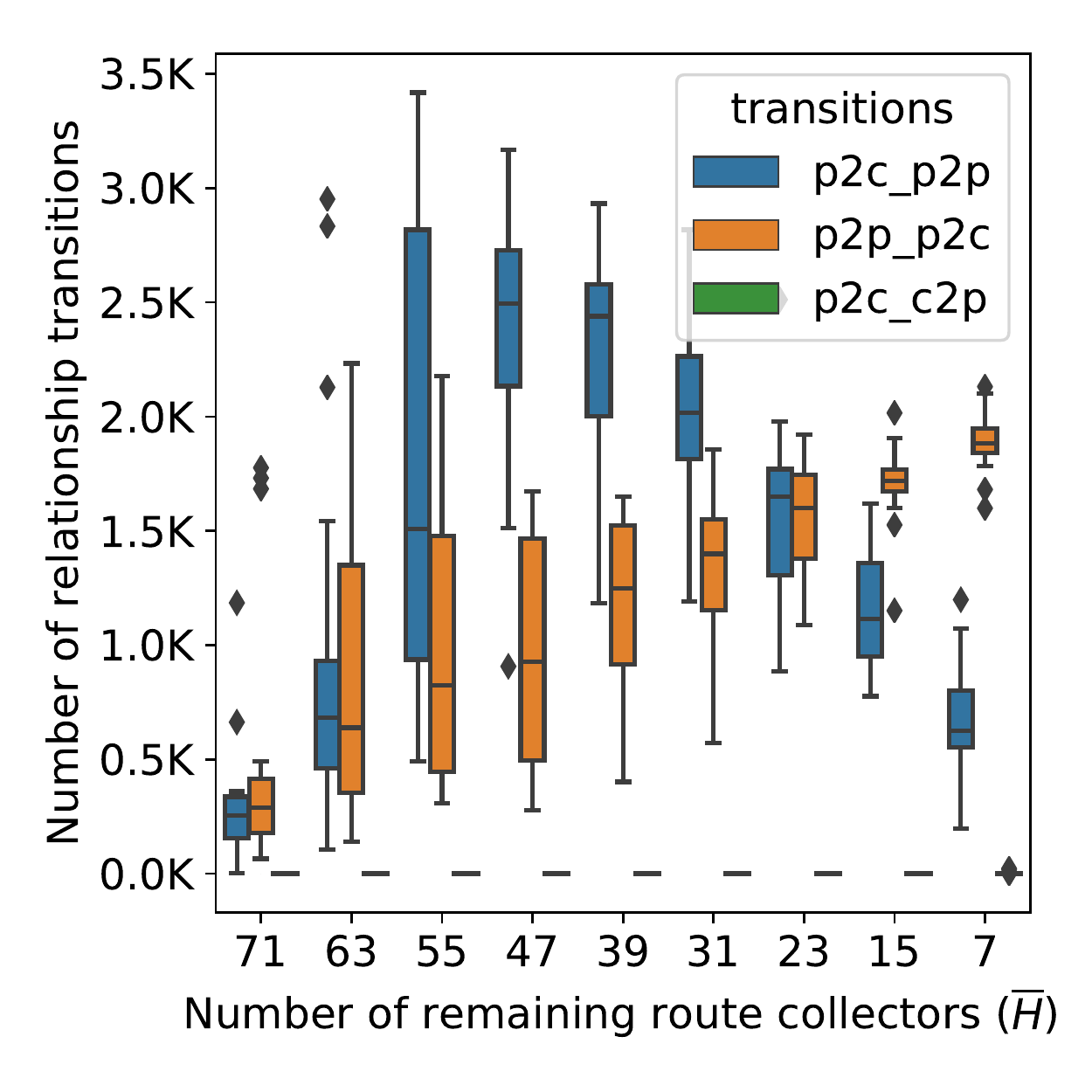}
		\subcaption{Relationship transitions as a function of $\overline{H}$ for \acrshort{mvp}.}
		\label{fig:mvp-transitions}
	\end{subfigure}%
	\begin{subfigure}[b]{.33\textwidth}
		\centering\captionsetup{width=.9\linewidth}
		\includegraphics[width=\textwidth]{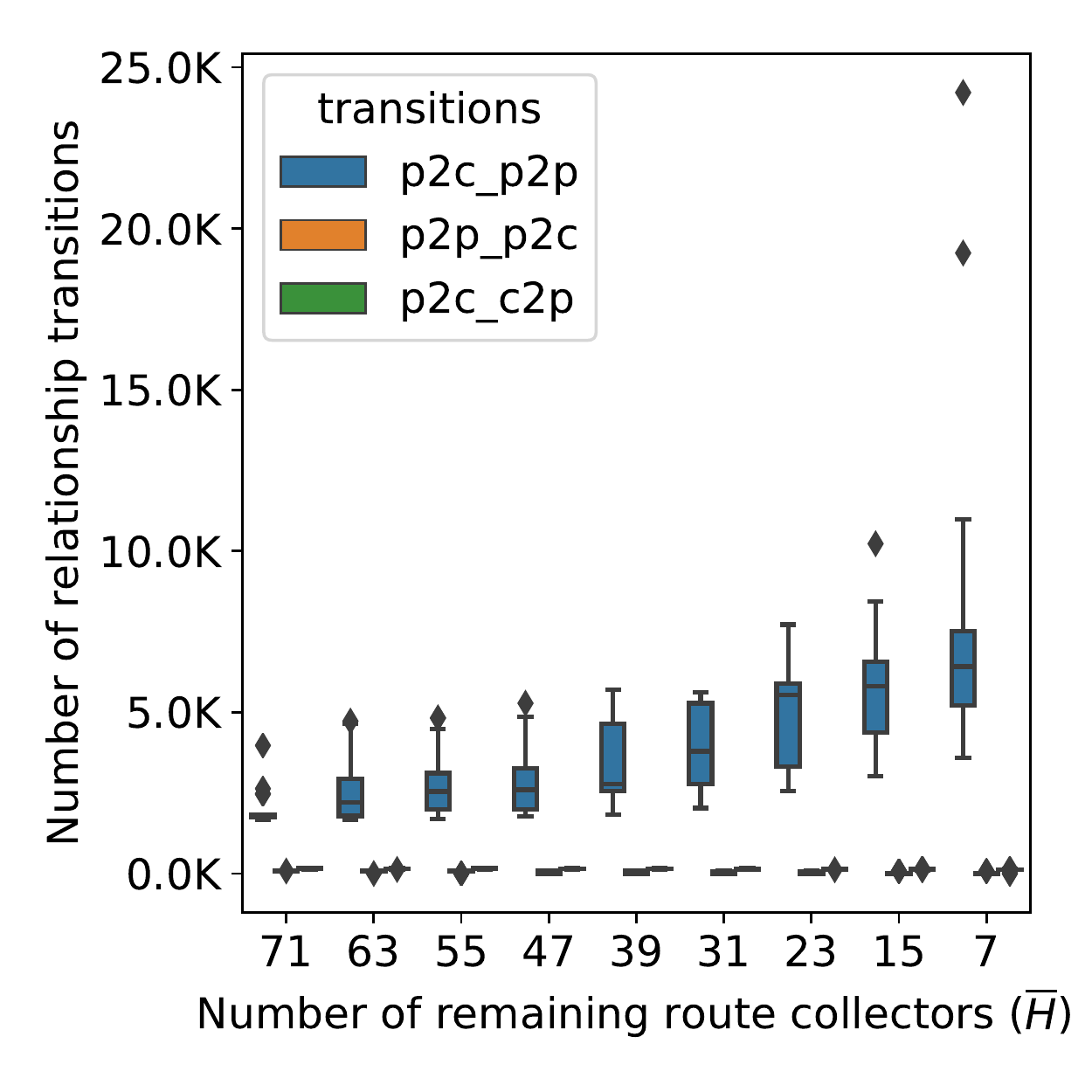}
		\subcaption{Relationship transitions as a function of $\overline{H}$ for \acrshort{ucla}.}
		\label{fig:ucla-transitions}
	\end{subfigure}%
	\begin{subfigure}[b]{.33\textwidth}
		\centering\captionsetup{width=.9\linewidth}
		\includegraphics[width=\textwidth]{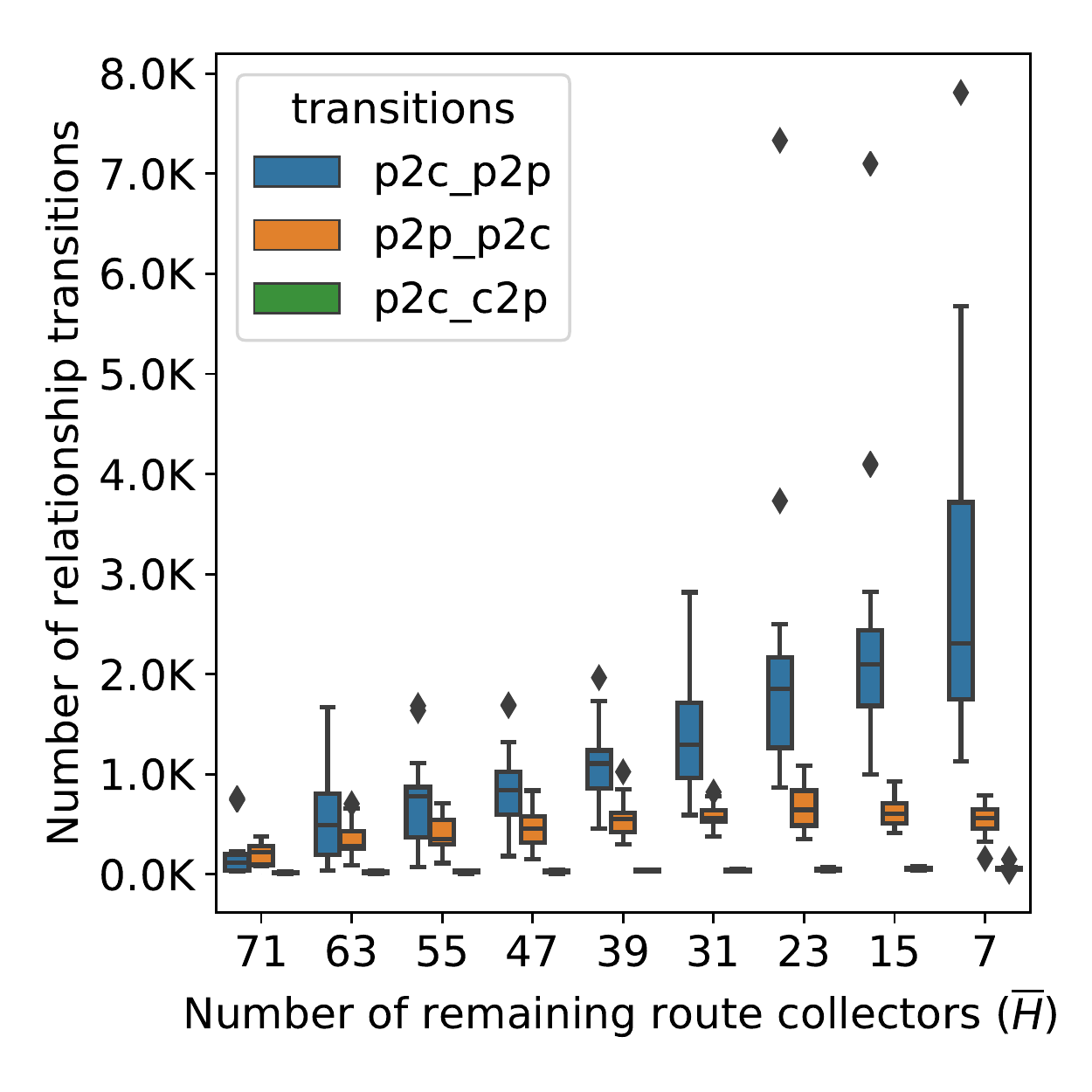}
		\subcaption{Relationship transitions as a function of $\overline{H}$ for \acrshort{asrank}.}
		\label{fig:asrank-transitions}
	\end{subfigure}
	\caption{Impact of reducing coverage.}
	\label{fig:transitions}
\end{figure*}

\textbf{Stability Results.}
Figure~\ref{fig:links-removed} and~\ref{fig:ASes-removed} show the increasing trend in the number of links and ASes removed as we progressively shrink $\overline{H}$. As $|\overline{H}|$ declines from 79 to 7 through multiple steps, the average loss of ASes rises from 0 to 2301 (5.6\%). 
However, the number of links experiences a dramatic drop, losing coverage of more than 38\% (on average) of the links observed by $I$. It is also important to realize that the number of links removed is not proportional to the number of excluded datasets because link coverage by different route collectors may overlap, especially at the top of the network hierarchy.

Figure~\ref{fig:gao-transitions} to~\ref{fig:asrank-transitions} show the number of relationship transitions by the four algorithms as a result of shrinking $\overline{H}$. At each step of shrinking $\overline{H}$, we measure the number of links whose inferred relationship is different under $I$ and $\overline{H}$. For instance, the curves labeled `p2p\_p2c' represent links whose inferred relationship changes from p2p to p2c with respect to $I$. In the case of \acrshort{ucla} (\ref{fig:ucla-transitions}), the curves of `p2c\_p2p' is trending upwards as more collectors are excluded and eventually reaches a mean of 7.8k transitions. In contrast, the curves of 'p2c\_c2p' and 'p2p\_p2c' stay mostly flat. Judging from the rightmost column, one is tempted to think that an average of 8k total transitions do not seem substantial with respect to the 128k links in $I$. But we note that it only requires the loss of coverage over 38\% of the links and the transitions affect 10\% of the remaining links, as shown in Figure~\ref{fig:links-removed}. We observe a similar surge in the `p2c\_p2p' curve of \acrshort{asrank} (\ref{fig:asrank-transitions}), with a smaller magnitude. Its `p2p\_p2c' curve experiences a gradual increase as $|\overline{H}|$ shrinks from 79 to 23 and witnesses a minor drop as the collector count decreases from 23 to 15. We also notice a small number of transitions from p2c to c2p. In contrast, the 'p2c\_p2p' curve of \acrshort{gao} (\ref{fig:gao-transitions}) follows the shrinking size of $\overline{H}$ and trends downward. What is surprising, however, is the large number of transitions at $|\overline{H}|=71$, caused by a minor drop in the number of route collectors. \acrshort{mvp} (\ref{fig:mvp-transitions}) seems more stable than the rest as measured by the number of transitions. The results support our speculation that these algorithms are unstable.

The observations illustrate the level of stability for different algorithms. While experimental results~\cite{luckie2013relationships} show the best-performing algorithms such as \acrshort{asrank} and \acrshort{ucla} achieve high precision and recall ($> 90\%$) in inference, they are sensitive to the number of route collectors available. In addition, the level of sensitivity may vary depending on the algorithm.

\subsection{Uncertainty in p2p Relationship Inference}
\label{sec:p2p-uncertainty}
It is hard to unambiguously infer p2p relationships because commonly used techniques, such as the valley-free constraint, are incapable of determining that a link must be p2p and completely eliminate the possibility of it being p2c~\cite{di2003computing, dimitropoulos2007relationships}. Consider a valid path with at least one AS link. If all links are c2p/p2c, there must exist a top provider adjacent to one or two customers under the valley-free constraint. The path remains valid if we classify a p2c link incident to the top provider as p2p. Conversely, given a valid path with a p2p link, replacing the p2p relationship with p2c or c2p will still leave the path valid. In general, this makes it more difficult to infer p2p than p2c relationships and the inference algorithms tend to leave the inference of p2p relationships as the last step after all p2c inferences are made.

\begin{figure}[htb!]
	\centering
	\includegraphics[width=0.9\linewidth]{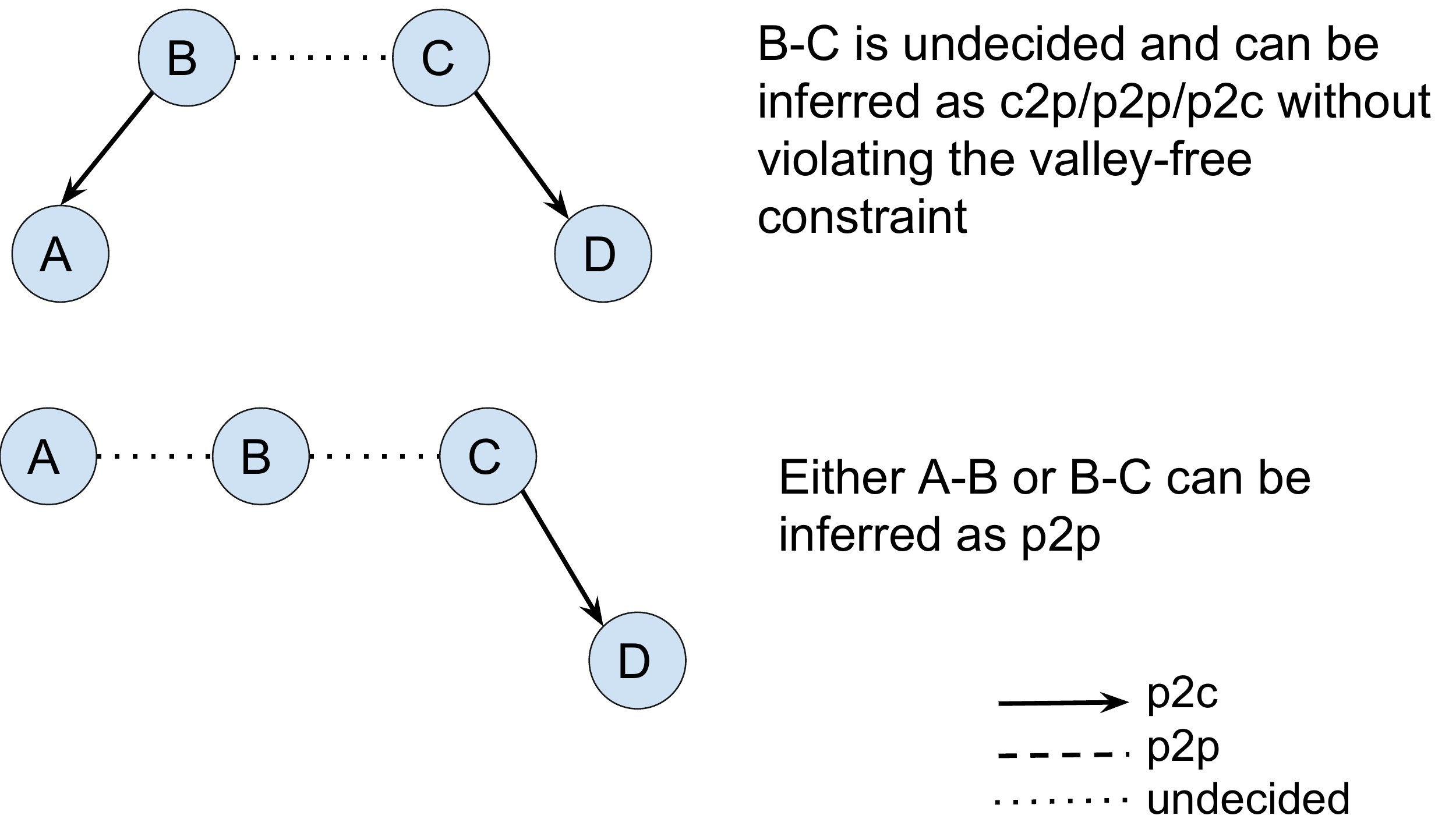}
	\caption{An example illustrating the challenge of inferring p2p relationships with certainty.}
	\label{fig:ucla-example}
\end{figure}

Figure~\ref{fig:ucla-example} is an example illustrating the challenge of inferring p2p relationship with certainty. Consider a path $A-B-C-D$. In the first scenario, suppose that \acrshort{ucla}, or any algorithm that prioritizes p2c/c2p inference over p2p~\cite{gao2001inferring, dimitropoulos2007relationships, luckie2013relationships}, is able to infer $C-D$ as p2c and $A-B$ as c2p. Algorithms typically infer the undecided link $B-C$ as p2p. But it could in fact be inferred as any one of the three relationships without violating the valley-free constraint. However, algorithms pick a single relationship, despite the uncertainty inherent to the scenario. Now consider another pathological scenario where the algorithm is only able to infer $C-D$ as p2c, leaving two consecutive links undecided. Inferring both $A-B$ and $B-C$ as p2p violates the valley-free constraint. Traditional algorithms aim to maintain the valley-free property and assign a single relationship, so they must speculate whether there exists a p2p link on this path and if yes, which of $A-B$ and $B-C$ should be inferred as p2p.

Prior algorithms have developed techniques to achieve high accuracy in p2p inference, but they are not as effective in handling such scenarios. For example, a common approach is to accurately infer as many p2c/c2p relationships as possible and the remaining links are then inferred as p2p\footnote{We focus our discussion on c2p/p2p/p2c because s2s relationships are uncommon.}. The accuracy of p2p inference now hinges on the performance of p2c/c2p inference. However, the lack of evidence to support p2c/c2p inference is not a guarantee that a link is p2p, as the above example shows. As another example, \acrshort{ucla} assumes that a provider wants to announce paths from its customers to its providers and the monitors at Tier-1 ASes should be able to reveal all the downstream p2c relationships over time~\cite{oliveira2010completeness}. A link hidden from all the Tier-1 ASes is less likely to be p2c, and thus should be inferred as p2p. Again, this rule only reduces rather than eliminates the possibility of incorrect p2p inference. The growing number of region-specific p2c relationships observable only below the provider AS~\cite{luckie2013relationships} makes the technique even less reliable. We conclude that we need a new strategy to express this inherent uncertainty in the inference results.

\subsection{Algorithmic Paradigms}
\label{sec:paradigm-limitation}
Broadly speaking, prior algorithms follow one of the two design paradigms: greedy~\cite{gao2001inferring,xia2004evaluation,subramanian2002characterizing,oliveira2010completeness,luckie2013relationships} and optimization-based~\cite{di2003computing,dimitropoulos2007relationships}. An algorithm's design paradigm is a deciding factor in its ability to identify the most probable relationship inference from multiple valid candidates. Using a collection of rules and heuristics, a greedy algorithm typically computes relationship inference over AS links in a sequential order. For a given link, its inference result might depend on another link's result computed in a previous step. For instance, suppose there is an AS path $v_{1}-v_{2}-v_{3}$ and the greedy algorithm infers $v_{1}-v_{2}$ as p2c in a previous step. It can apply the valley-free constraint to infer $v_{2}-v_{3}$ as p2c next.

A greedy approach to AS relationship inference has a critical limitation: it does not support backtracking and branching. When making inference on an AS link, the algorithm must commit to one of the feasible relationships as governed by the rules and heuristics. If a mistake is made, a greedy algorithm cannot recover from it and the error can impact the inference results for the remaining links. Returning to the example above, we suppose that the algorithm has already inferred $v_{1}-v_{2}$ as p2c and $v_{2}-v_{3}$ as p2p, before it encounters the path $v_{1}-v_{2}-v_{3}$. Clearly the path now violates the valley-free constraint. When the p2p inference was made on $v_{2}-v_{3}$, however, it was possible that both p2p and p2c were feasible according to the valley-free constraint. Then a heuristic was used to break the tie and gave precedence to p2p over p2c. The violation could have been avoided if the algorithm had chosen p2c instead. Unfortunately, greedy algorithms lacks the capability to maintain the uncertainty of relationship inference across steps.

In contrast to the greedy approach, an optimization-based algorithm explores all feasible inferences simultaneously for all links. It usually encodes the heuristics and rules as constraints and optimize for a certain objective. For example, DPP~\cite{di2003computing} uses boolean variables to represent the orientation of links (i.e., p2c v.s. c2p) and a collection of 2-literal disjunctive clauses to express the valley-free constraint. The objective is set to minimize the number of unsatisfiable clauses, or equivalently, the number of invalid paths. But~\cite{dimitropoulos2007relationships} identifies the discrepancy that maximizing the number of valley-free paths does not necessarily translate to higher inference accuracy. In general, the objective of optimization-based algorithms is not fully aligned with the ideal objective to maximize the accuracy because the true relationships are not known and it is difficult to quantify the difference between the two objectives. It is this discrepancy in the two objectives that affects our level of confidence in the inference output and leads to uncertainty.

The main takeaway is that a new paradigm should be able to overcome the limitations of greedy and optimization-based algorithms.

\subsection{Unreliable Assumptions and Rules}
\label{sec:flaky}
In addition to the valley-free constraint, inference algorithms must rely on additional assumptions to derive rules and heuristics. These assumptions vary in terms of reliability but prior algorithms fail to differentiate between them in terms of confidence level when making inferences. 

For example,~\cite{oliveira2010completeness} and~\cite{luckie2013relationships} assume there exists a clique of Tier-1 providers interconnected through p2p links. It is easy to obtain a list of Tier-1 providers and the rule derived from the assumption achieves near-perfect inference accuracy when validated using a partial ground truth dataset~\cite{luckie2013relationships}. However, these algorithms also use assumptions that depend on a predefined parameter or ``magic constant''. For example, \acrshort{gao} algorithm compares the ratio of node degrees against a constant to determine the provider and customer roles for adjacent ASes. Links that are categorized using this parameter are not marked any differently than those using near-perfect rules. As another example,~\cite{luckie2013relationships} assumes that vantage points reporting routes to less than 2.5\% of ASes are not announcing provider routes to the route collector. It is unclear why the $2.5$\% threshold is appropriate and how a different value can affect the accuracy. We believe unreliable assumptions can introduce uncertainty into the inference process and the labels generated by the corresponding rules should convey the uncertainty of the results.

\subsection{Potential Benefits of Exposing Inference Uncertainty}
The results of AS relationship inference have supported numerous research works~\cite{huang2007can, shah2011rumors, mahajan2002understanding, calder2013mapping, luckie2014challenges}. Exposing the uncertainty in inference opens up the opportunity to improve the robustness of their findings. We use two prior works to illustrate the potential benefits. \cite{huang2007can} used AS relationships to evaluate the impact of peer-assisted video-on-demand (VoD) on ISPs. The authors showed that ISP-friendly peer-assisted VoD achieved 50\% savings compared to solutions with no P2P. They also stated that the inference by CAIDA had been conservative because ISPs were unwilling to share their sibling and peering relationships. It is possible that quantifying the uncertainty in inference helps develop a bound rather than a single value on the estimated savings. As another example, \cite{mahajan2002understanding} relied on AS relationships to understand BGP misconfigurations. The export policies, derived from the relationships, enabled the authors to identify export misconfigurations. The authors acknowledged that incorrect inference could lead to failure in identifying misconfigurations or mistake legitimate configurations as erroneous. Based on the uncertainty in inference, it might be possible to estimate the uncertainty in identifying misconfigurations.
	\section{Towards Probabilistic AS Relationship Inference}
\label{sec:rationale}
As we demonstrate in Section~\S\ref{sec:uncertainty}, uncertainty is inherent in AS relationship inference but largely ignored by prior algorithms. To bridge this gap, we propose a new framework that explicitly reflects uncertainty in the inference results. An exemplary algorithm is presented in~\S\ref{sec:pari}.

We start by deriving the design requirements based on the insights from~\S\ref{sec:uncertainty}. First, when it comes to single-label (deterministic) inference, it should not rely on heuristics whose output may vary depending on the amount of information available, such as heuristics based on node or transit degrees ((\S\ref{sec:stability})).  This allows us to minimize the uncertainty caused by an incomplete view of the AS topology. Second, it should expose uncertainty by associating more than one type of relationship with an AS link if necessary~\footnote{The \acrshort{gao} algorithm may infer multiple relationships on one link but it seems to be an artifact of not enforcing mutual exclusion in the algorithm design. In fact, the total number of inferences exceed the number of links in the topology in the evaluation results~\cite{gao2001inferring}.} (\S\ref{sec:p2p-uncertainty}). Third, it must differentiate unreliable assumptions and heuristics from the more trustworthy rules, such as the valley-free constraint (\S\ref{sec:flaky}). Fourth, it should mitigate the limitations of greedy and optimization-based paradigms (\S\ref{sec:paradigm-limitation}).

Our framework consists of two major steps. After extracting a set of AS paths from the dataset, we identify potentially erroneous AS paths and remove or sanitize them. The remaining AS paths, collectively forming the AS topology, are then passed to the principle-based inference step where we attempt to apply~\textit{principles}, such as the valley-free constraint and other reliable rules, to deterministically infer one of the three possible relationships for each link in the topology (\acrlong{principle}). If the principles manage to eliminate all but one relationship, the inference is complete for this link; otherwise, we leave it to the next step. Note that \acrlong{principle} is intentionally conservative and incorporates rules such that it commits to an inferred relationship only if the confidence level is very high. Based on potentially less reliable assumptions and heuristics, the final step (\acrlong{probabilistic}) performs probabilistic inference and computes a \textit{tag}, that represents our confidence in the inference, for each remaining link. The tag can be numerical or categorical. The separation of principle-based and probabilistic inference allows us to expose the level of uncertainty associated with each link. One might argue that our framework also suffers from drawbacks of the greedy algorithms since misclassifications made in \acrlong{principle} can impact the tags calculated in \acrlong{probabilistic}. But the conservative selection of rules in \acrlong{principle} prioritizes accuracy over coverage and thus minimize such mistakes. We defer describing how our framework also addresses limitations of the optimization-based approach to the concrete implementation in \S\ref{sec:pari} where we exploit a partial ground truth dataset and formulate the objective to maximize inference accuracy directly.

\subsection{A Flexible Framework}
The proposed framework does not prescribe the specific rules to be
used in \acrlong{principle} and \acrshort{probabilistic}. We
demonstrate the framework's flexibility by illustrating the very
diverse choices that can be made in the two steps.

\textbf{Principle selection for \acrlong{principle}.} To attain high
certainty, \acrlong{principle} could for example rely on the
valley-free constraint.  It can also incorporate external knowledge
such as the set of all Tier-1 ASes.

\textbf{Heuristic selection for \acrlong{probabilistic}.}
\acrlong{probabilistic} is the major enabler of probabilistic
inference, allowing not only a wide variety of heuristics but also
different tag formats to express the confidence in the inference
output.

First, the output of \acrlong{probabilistic} can be a categorical tag
indicating ``high'' or ``low'' confidence.  For example, building on
prior work, recall that in Figure~\ref{fig:ucla-example}, $B$ and $C$
can be inferred as any one of the three relationships without
violating the valley-free constraint. Now suppose we further observe
that link $B-C$ is not seen in any AS path announced to $B$ or $C$'s
providers. It is an indication that $B-C$ is more likely to be
p2p. However, the heuristic exploits a ``negative observation''; it is
grounded in the lack of paths that the vantage points expect to
observe and we have seen impact of incomplete coverage in
\S\ref{sec:stability}. Thus, labeled links can receive a ``high'' or
``low'' confidence tag depending on whether they were classified based
on the valley free constraint or based on negative observation
heuristic.  Optionally, we can also document the reason for the low
confidence tag, $B-C$ would be labeled as (p2p, ``low'', ``negative
observation'').  We note that \cite{oliveira2010completeness,
  luckie2013relationships} explore similar ``negative observations''.

As another example, the tag could be numeric, e.g., a scalar.  
Recall from \S\ref{sec:flaky} that \acrshort{gao}
algorithm compares the ratio of node degrees against a constant to
determine the provider and customer roles for adjacent ASes. The
intuition is that the extent to which the ratio deviates from 1 is an
indication of how likely that one of the ASes is transiting traffic
for the other. Thus, we may simply use this ratio as the numerical tag
along with the c2p/p2c label. As for links inferred as p2p, we may set
the tag value to be inversely proportional to the difference in
transit/node degree of the end points.

\begin{figure}[htb!]
	\centering
	\includegraphics[width=0.9\linewidth]{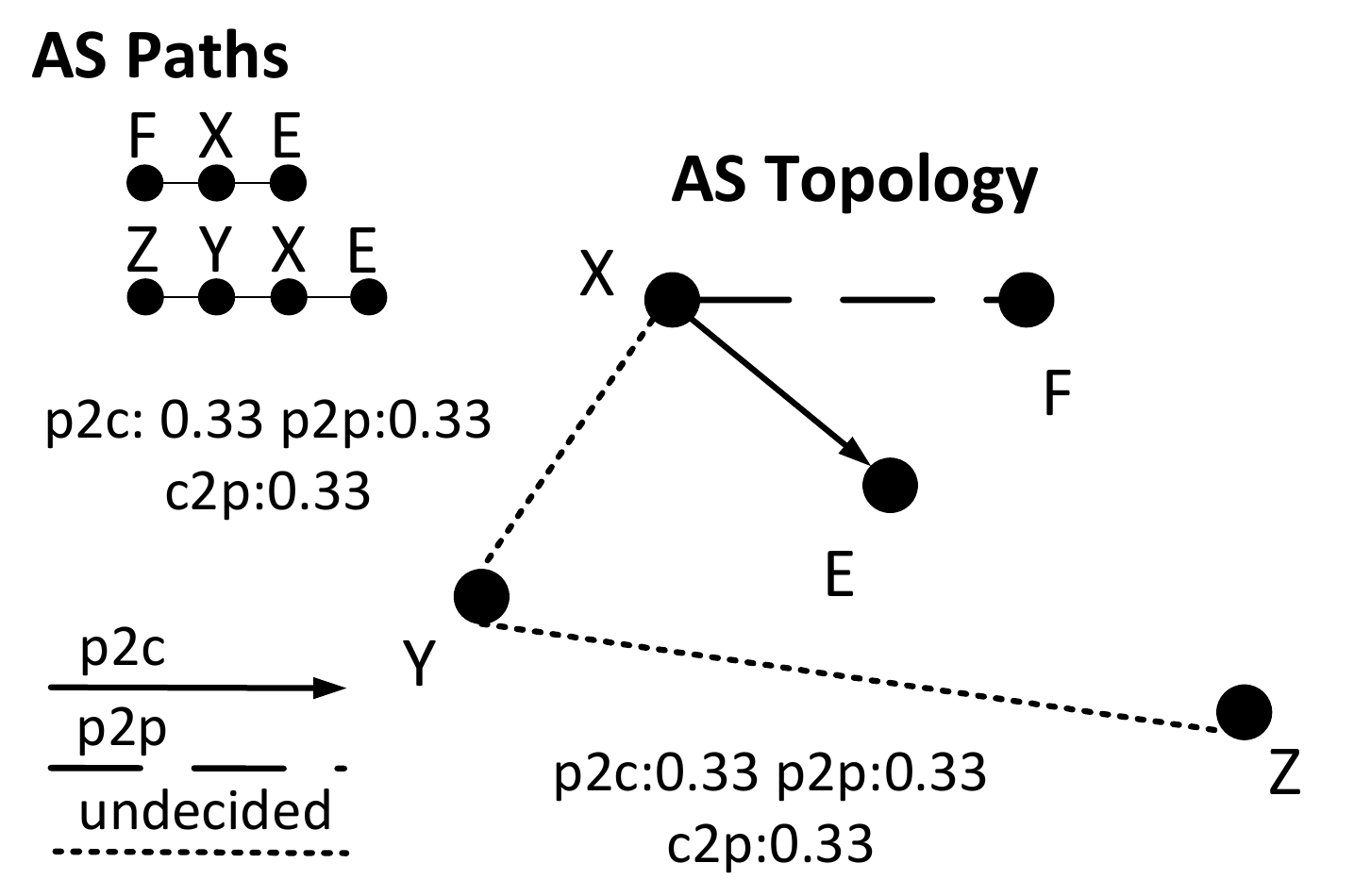}
	\caption{An example to implement the probabilistic AS relationship inference framework. \acrlong{principle} makes high-confidence inferences over links $X-E$ and $X-F$. \acrlong{probabilistic} exposes inferences with high uncertainty over links $X-Y$ and $Y-Z$ through tags, which are score vectors in this example.}
	\label{fig:algo-diag}
\end{figure}

We can also introduce richer tags and new heuristics. Consider a
heuristic that outputs as a tag a \textit{vector of scores}, where
each score represents our confidence in inferring the link as one of
the three relationships. For example, it may assign equal score values
to all feasible relationships according to the valley-free
constraint. We provide an example in Figure~\ref{fig:algo-diag} where
the AS topology is formed by two paths. Suppose that $X-F$ and $X-E$
are inferred as p2p and p2c respectively in \acrlong{principle}. At
this point, the other 2 links have two or more feasible relationships
under the chosen principles. In this example, we use the
aforementioned heuristic that assigns equal score values to all
feasible relationships. Observe that if we infer $X-Y$ as any one of
c2p/p2p/p2c, there always exists an inference of $Y-Z$ such that all
paths are valley-free. Thus, the heuristic assigns $\frac{1}{3}$ to
all elements in the score vector of $X-Y$.

\section{Probabilistic AS Relationship Inference Algorithm}
\label{sec:pari}
In this section, we describe the Probabilistic AS Relationship Inference (\acrshort{pari}) algorithm that leverages our framework from~\S\ref{sec:rationale}.~\S\ref{sec:preprocess} to~\S\ref{sec:prob-inf} describe in detail the implementation of \acrlong{principle} and \acrshort{probabilistic} of the framework.

\subsection{Preprocessing AS Paths}
\label{sec:preprocess}
Similar to the filtering and sanitizing step described in~\cite{luckie2013relationships}, \acrshort{pari} starts by preprocessing the set of AS paths. It first removes AS paths with loops and paths where two Tier-1 ASes are separated by one or more non-Tier-1 ASes. Both of these path types are signs of path poisoning usage. It also removes paths with unassigned ASes. For each path, if it identifies an AS used to operate IXP route servers, \acrshort{pari} removes it and establishes a link between its preceding and following neighbors.

\subsection{Principle-based Inference}
\label{sec:principle-inf}
\acrshort{pari}'s principle-based inference adopts as principles two rules commonly used by the recent inference algorithms~\cite{oliveira2010completeness,luckie2013relationships}. For the first principle, \acrshort{pari} leverages the observations that there exists a clique of Tier-1 ASes interconnected with p2p links at the top of the hierarchy and that the AS members of the clique are known. We can either identify the clique through preprocessing~\cite{luckie2013relationships}  or obtain the list of top-tiers from public sources~\cite{oliveira2010completeness}. The second principle imposes the constraint that the relationships assigned to links along an AS path must be valley-free.

After extracting a collection of valid AS paths from the dataset, \acrshort{pari} infers the links connecting a pair of Tier-1 ASes as p2p based on the first principle. Then, \acrshort{pari} infers a set of p2c links, with high confidence, following the second principle. The inference of p2c links has two phases. Phase~\Rmnum{1} exploits the knowledge of Tier-1 ASes. Given an AS path $p_j: v_1 - v_2 - \cdots - v_h - v_{h+1} - \cdots - v_{m} \in Q$, \acrshort{pari} searches for the last Tier-1 AS starting from $v_1$. Suppose $v_{h}$ is the last Tier-1 AS along the path, we can deduce that $v_{h}-v_{h+1}$ is of type p2p or p2c because a Tier-1 AS has no provider. Moreover, the type of every link on the path segment $v_{h+1} \rightsquigarrow v_{m}$ must be p2c by the valley-free principle. PARI applies this procedure on each path and ends up with a set of p2c inferences. An idea similar to Phase~\Rmnum{1} of p2c inference was proposed in~\cite{oliveira2010completeness}. However, \acrshort{pari} takes it one step further and introduces a Phase~\Rmnum{2} of p2c inference. It takes advantage of the outputs from Phase~\Rmnum{1} and infers as many p2c links as possible by repeatedly iterating through the list of paths. Within each iteration, it applies the valley-free principle based on links that have been inferred in the previous iterations and makes p2c inference on undecided links accordingly. For example, consider another AS path $p_i: v_{h+1} - v_{h+2} - v_{w} \in Q$ where none of the nodes is Tier-1 AS. While Phase~\Rmnum{1} cannot infer p2c links on $p_i$ directly, one of its links, $v_{h+1} - v_{h+2}$, is inferred as p2c after $p_j$ is processed. The first iteration of Phase~\Rmnum{2} is able to apply the valley-free principle and infer $v_{h+2} - v_{w}$ as p2c. The succeeding iterations can now leverage the inference result on $v_{h+2} - v_{w}$. Phase~\Rmnum{2} terminates when no new p2c links are produced in the last iteration.

Conflicts might arise from the procedure above. Specifically, a link $u - v$ might be inferred as p2c in one path but c2p in another path. The problem can be attributed to the observation of s2s links, where both end points can export their own routes and the routes of their customers, providers or peers~\cite{gao2001inferring}. A study~\cite{dimitropoulos2007relationships} has shown that 31 out of 3724 verified relationships are s2s according to its survey data. \acrshort{pari} labels these links as ``conflict'' and leaves final classification of the link as either p2c or c2p to the probabilistic inference step.

The principle-based inference step is summarized in Algorithm~\ref{algo:principle}.
\begin{algorithm}
	\caption{Principle-based Inference}
	\label{algo:principle}
	\begin{algorithmic}[1]
		\State Infer p2p links between Tier-1 ASes
		\State Infer p2c links following Tier-1 ASes \Comment{Phase~\Rmnum{1}}
		\State Gather all inference results into link set $\mathcal{J}$
		\Repeat \Comment{Phase~\Rmnum{2}}
		\State Scan through the path list to infer more p2c links using $\mathcal{J}$
		\State Update $\mathcal{J}$ to incorporate the new p2c links
		\Until{No new p2c links inferred in the last iteration}
		\State Label conflicting p2c inferences
	\end{algorithmic}
\end{algorithm}

\subsection{Probabilistic Inference}
\label{sec:prob-inf}
In the last step, \acrshort{pari} performs probabilistic inference over the remaining undecided links. It starts with a procedure to compute a \textit{share vector}, denoted by $S$, for every undecided link. The vector measures the number of valley-free inferences over the entire topology, under the condition that the link is inferred as one of the relationships. Then we refine the procedure with a relaxation step to handle the case where no valley-free inference can be found. Finally, \acrshort{pari} outputs the score as the conditional probability that the type of relationship of a link is $r$ given its share vector $s$. We choose \acrfull{mlr} and leverage a partial ground truth dataset for learning the probabilistic model. 

\subsubsection{Computing Shares}
\label{sec:comp-sv}
\begin{figure}[htb!]
	\centering
	\includegraphics[width=0.8\linewidth]{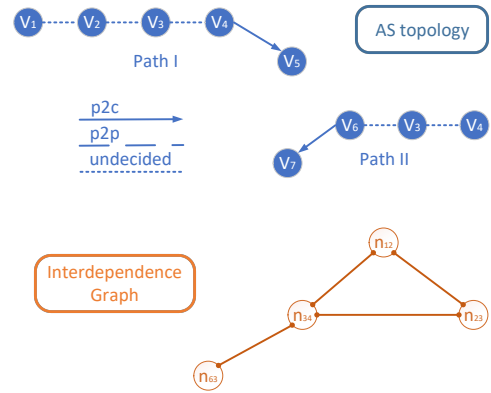}
	\caption{
		An AS topology with undecided links and its interdependence graph ($\mathcal{G}$).
	}
	\label{fig:interdependence-eg}
\end{figure}

By the valley-free principle, the inferred types of relationship of undecided links found within the same AS path are interdependent. For example, Figure~\ref{fig:interdependence-eg} shows an AS topology with links $v_{4} - v_{5}$ and $v_{6} - v_{7}$ inferred as p2c by principle-based inference while the other links are undecided. In Path~\Rmnum{1}, if we infer any one of the undecided links as p2p, the other two links must be inferred as p2c or c2p; otherwise, it violates the valley-free principle.

\begin{table}[htb!]
	\centering
	\caption{All valley-free Inferences for the topology in Figure~\ref{fig:interdependence-eg}.}
	\label{tab:valley-free-infer}
	\begin{tabular}{@{}llll@{}}
		\toprule
		$n_{12}$   & $n_{23}$  & $n_{34}$ & $n_{36}$ \\
		\midrule
		p2c	&	p2c	&	p2c	&	p2c/p2p/c2p\\
		p2p	&	p2c	&	p2c	&	p2c/p2p/c2p\\
		c2p	&	p2c	&	p2c	&	p2c/p2p/c2p\\
		c2p	&	p2p	&	p2c	&	p2c/p2p/c2p\\
		c2p	&	c2p	&	p2p	&	p2c\\
		c2p	&	c2p	&	c2p	&	p2c\\
		\bottomrule
	\end{tabular}
\end{table}

To capture the interdependence imposed by valley-free principle, we introduce a graph called interdependence graph ($\mathcal{G}$). Each undecided AS link corresponds to a node in $\mathcal{G}$. For each pair of nodes in the topology, we connect them with an arc iff there exists at least one AS path that contains both AS links. For example, the two paths in Figure~\ref{fig:interdependence-eg} leads to four nodes $n_{12}: v_{1} - v_{2}$, $n_{23}: v_{2} - v_{3}$, $n_{34}: v_{3} - v_{4}$, and $n_{36}: v_{3} - v_{6}$. An arc is created for a pair of nodes if they appear in the same path. Note that the interdependence relationship can span across different AS paths if they share the same undecided links. In the example, $v_{3} - v_{4}$ is the common link shared by Path~\Rmnum{1}\&\Rmnum{2}.

Now we associate each node $n_{i}$ in $\mathcal{G}$ with three Boolean variables $x_{i}$, $y_{i}$, and $z_{i}$ and use them to denote c2p, p2p, and p2c relationships respectively. The valley-free principle can be expressed as a Boolean formula over these variables. Returning to the example in Figure~\ref{fig:interdependence-eg}, we can use $y_{23} \Rightarrow z_{34}$ to express the following constraint: if node $n_{23}$ is inferred as p2p, node $n_{34}$ must be p2c. By solving for a satisfying assignment of the formula, we can identify a valid valley-free inference over the undecided links. Moreover, the valley-free principle is typically insufficient to produce a unique satisfying assignment. For instance, if we assume that Path~\Rmnum{1}\&\Rmnum{2} in Figure~\ref{fig:interdependence-eg} are the only paths containing the four undecided links, then these nodes form a single connected component in $\mathcal{G}$ and their inference are independent from other nodes in $\mathcal{G}$. Each satisfying assignment matches one of the valley-free inferences in Table~\ref{tab:valley-free-infer}.

We use $C_{j}$ to label the connected components in the $\mathcal{G}$ and $F_{E}(C_{j})$ to denote the number of unique valley-free inferences over $C_{j}$ given that the predicate $E$ evaluates to true. The subscript is dropped when no predicate is specified. Returning to Figure~\ref{fig:interdependence-eg}, the number of valley-free inferences is $F(C_{j})=14$ and the number of valley-free inferences given that $n_{12}$ is inferred as c2p is $F_{x_{12}}(C_{j})=8$. We express the share associated with $n_{i}$ and relationship $R$ as a normalized count
\begin{equation}
	S_{i}^{R} = \frac{F_{R_{i}}(C_{j})}{F(C_{j})}
\end{equation}
where $R \in \{x, y, z\}$ and $C_{j}$ is the connected component in which $n_{i}$ resides. In the running example, the share associated with $n_{12}$ and c2p is $S_{12}^{x}=\frac{8}{14}$. The computation of $F_{E}(C_{j})$ can be reduced to a propositional model counting problem~\cite{biere2009handbook}, or \#SAT.

The definition above makes two assumptions. First, it enforces that the correct inference for the connected component is valley-free. It is justified because valley-free violation is uncommon~\cite{giotsas2012valley} and the valley-free principle is fundamental. We will describe how \acrshort{pari} handles the case where no valley-free inference can be found in~\S\ref{sec:relaxation}. Second, it gives all possible valley-free inferences the same weight. We believe it is a reasonable assumption as it simplifies the computation of shares and leave part of the complexity of deciding the weights to the final step---computing probability estimates.

\subsubsection{Relaxation for Unsatisfiable Connected Components}
\label{sec:relaxation}
Some connected components might have no valley-free inference due to two reasons.
First,~\cite{giotsas2012valley} identified valley paths using BGP Community data. If $C_{j}$ contains nodes representing AS links from these paths, the algorithm is unable to find valley-free inference. Second, incorrect inferences are made during the principle-based inference step and propagate to the probabilistic inference step.

Inspired by the idea of maximizing the number of satisfiable clauses~\cite{dimitropoulos2007relationships, di2003computing}, we propose a relaxation method to mitigate these problems. As mentioned in~\S\ref{sec:prob-inf}, the valley-free constraint imposed over $C_{j}$ can be expressed as a Boolean formula. Our approach first leverages a MAX-SAT solver to determine an assignment maximizing the number of clauses that can be made true. For each clause that expresses an interdependence between AS links but is unsatisfied by the assignment, we remove the corresponding arc from $C_{j}$. At the end of this process, $C_{j}$ is broken into one or more connected components. The same process is applied to each resulting component recursively if it remains unsatisfiable. Instead of ignoring the unsatisfied clauses, as in~\cite{dimitropoulos2007relationships, di2003computing}, our method removes the interdependence that leads to these clauses.

Once an arc is removed from a component, the interdependence between two AS links, which must have appeared in the same AS path, is lost. The goal is to preserve as many interdependence relationships as possible. It is challenging to incorporate the existence of valley paths in \acrshort{pari} as the valley-free principle is fundamental in our approach to deriving the shares.

Algorithm~\ref{algo:comp-sv} summarizes the procedure of computing share vectors in \S\ref{sec:comp-sv} and \S\ref{sec:relaxation}. It begins by identifying a set of connected components, $\mathcal{U}$, from the interdependence graph. Then for each component it calculates the number of valley-free inferences $F(C_{j})$. If the value is 0, meaning the associated Boolean formula is unsatisfiable, it dismantles it into a set of smaller, satisfiable connected components, and merge them back to the group. Now that all connected components in the set are satisfiable, the nested for loop (line~\ref{lst:line:process-cj}) proceeds to calculate the share vectors for each link in every connected component, which is parallelizable because no dependency spans across multiple components. Our evaluation on real datasets show that the while loop (line~\ref{lst:line:relaxation}) terminates within two iterations and only a small percentage of links are removed due to splitting the connected components.

\begin{algorithm}
	\caption{Computing Share Vectors}
	\label{algo:comp-sv}
	\begin{algorithmic}[1]		
		\State Identify a set of connected components $\mathcal{U}=\{C_{j}\}$ from the interdependence graph $\mathcal{G}$~\label{lst:line:components}
		\While{$\exists C_{j} \in \mathcal{U}$ s.t. $F(C_{j})$ = 0}~\label{lst:line:relaxation}
			\State Remove $C_{j}$ from $\mathcal{U}$
			\State Compute a set of satisfiable connected components $\{C_{j}^\prime\}$ from $C_{j}$ using relaxation.
			\State Merge $\{C_{j}^\prime\}$ back to $\mathcal{U}$
		\EndWhile
		
		\ForAll{$C_{j} \in \mathcal{U}$} \label{lst:line:process-cj}
			\ForAll{$n_{i} \in C_{j}$}
				\State Compute $F_{R_{i}}(C_{j})$ and $S_{i}^{R}$ where ${R} \in \{x, y, z\}$
			\EndFor
		\EndFor
	\end{algorithmic}
\end{algorithm}

\subsubsection{Computing Probability Estimates}
\label{sec:comp-pe}
Finally, we apply \acrshort{mlr} to model the conditional probability $Pr(R=r|S=s)$ using the elements of share vector as features. We decide to use \acrshort{mlr} because it outputs not only the predicted type of relationship, but also the probability estimates. The ground truth dataset is partitioned into the training and test sets. \acrshort{mlr} assumes that $Pr(R=r|S=s)=f(s;\theta)$ for some function $f$ parameterized by $\theta$. Intuitively, it aims to estimate $\theta^\star$, the value of $\theta$ that is most consistent with the true relationship from the training set. The test set will be used to evaluate the model's performance. Now consider an unlabeled link beyond the partial ground truth dataset. Based on its score vector $s^\prime$, the trained model outputs $f(s^\prime;\theta^\star)$ to estimate the probability that its type of relationship is $r$.

Importantly, \acrshort{mlr} assumes that the relationship of undecided links are independent; but this is not true because the valley-free constraint induces interdependence across them. Markov Logic Network~\cite{richardson2006markov} (MLN) allows us to explicitly express the valley-free constraint using first-order logic. But our experience with an MLN inference engine (i.e., Tuffy~\cite{niu2011tuffy}), indicates that it faces scalability issues when applied to the AS-level Internet topology. Therefore, we consider the share vectors as a reasonable compromise and leaves the search of a more scalable MLN-based solution as future work.

	\section{PARI Evaluation}
\label{sec:evaluation}
In this section, we evaluate the performance of \acrshort{pari}. First, we show the results of the preprocessing step and introduce the set of tools used for the implementation. Then, we present the evaluation results for \acrlong{principle} and \acrshort{probabilistic}. The details of the BGP data and the partial ground truth data are described in \S\ref{sec:background}.

We rely on public BGP datasets in this evaluation. \acrshort{pari}'s preprocessing step extracts 7,153,208 out of 13,482,724 unique AS paths from these datasets. The result is a topology of 41,127 ASes and 130,076 links~\footnote{The topology after preprocessing has more links than the topology formed by all route collectors from $I$ in \S\ref{sec:background} because \acrshort{pari} introduces new links between the peering participants at the IXPs when the ASes that are serving as route servers are removed from AS paths.}.

In \acrshort{pari}'s probabilistic inference, we use sharpSAT~\cite{thurley2006sharpsat} for counting the number of valid assignments over each connected component and Open-WBO~\cite{martins2014open} to determine the maximum number of satisfiable clauses for unsatisfiable components.

\subsection{Principle-based Inference Results}
We evaluate the principle-based inference step in terms of accuracy and stability. First, we show that PARI achieves higher accuracy than prior algorithms when it infers a single relationship on an AS link. We then turn to stability evaluation and analyze how the principle-based inference step helps address the issue in \S\ref{sec:stability}.

\subsubsection{Accuracy}
A set of 22 Tier-1 ASes is retrieved from~\cite{tier-one}\footnote{For consistency, we verify that they match the Tier-1 ASes identified in \cite{luckie2013relationships}, the source of the ground truth dataset.}. \acrshort{pari} infers 151 p2p links across them, among which 131 are labeled as p2p in the ground truth dataset. One Tier-1 p2p link is not present in any of the paths in the input, so it cannot be inferred by \acrshort{pari}.
Therefore, \acrshort{pari} correctly infers all validated p2p links in the clique of Tier-1 ASes.

The principle-based inference step infers 78,689 c2p/p2c links and 383 s2s links, among which 26,637 links are found in the ground truth dataset. Its confusion matrix is shown in Table~\ref{tab:principle-results}. Observe that \acrshort{pari} correctly infers 26,339 out of 26,389 validated p2c links but it misclassifies 26 p2c links as c2p. 64 p2c links are inferred as both p2c and c2p. As described in \S\ref{sec:principle-inf}, they represent a very small group of links experiencing conflicting inference results; each of them appears in multiple AS paths and applying the principles does not lead to the same relationship across all paths. The principle-based inference step does not infer p2p links, but it misclassifies a total of 208 p2p links as p2c, c2p, or conflict.

\begin{table}[!htb]
	\centering
	\caption{Confusion matrix of principle-based inference step. Conflict links are inferred as both c2p and p2c.}
	\label{tab:principle-results}
	\begin{tabular}{@{}ccccc@{}}
		\toprule
		& \multicolumn{3}{c}{Inferred relationship} &	\\
		\cmidrule{2-4}
		True relationship	&	p2c	&	c2p	&	conflict	&	Total\\
		\midrule
		p2c	&	\textbf{26339}	&	26	&	64	&	26429\\
		p2p	&	110	&	93	&	5	&	208\\
		\bottomrule
	\end{tabular}
\end{table}

The accuracy of the principle-based inference step is 98.9\%. Over the same set of validated links, \acrshort{asrank} and \acrshort{ucla} achieve accuracies of 96.9\% and 96.3\%, respectively. Note that they are consistent with the overall performance of \acrshort{asrank} and \acrshort{ucla} when applied to all validated links in the topology (97\% and 95.5\%). It means the subset of links are not deliberately chosen to favor~\acrshort{pari}. To conclude, the principle-based step prioritizes reliability over coverage, achieving higher accuracy than \acrshort{asrank} and \acrshort{ucla} on a subset of links.

\subsubsection{Stability}
\begin{figure}[!htb]
	\centering
	\includegraphics[width=1.0\linewidth]{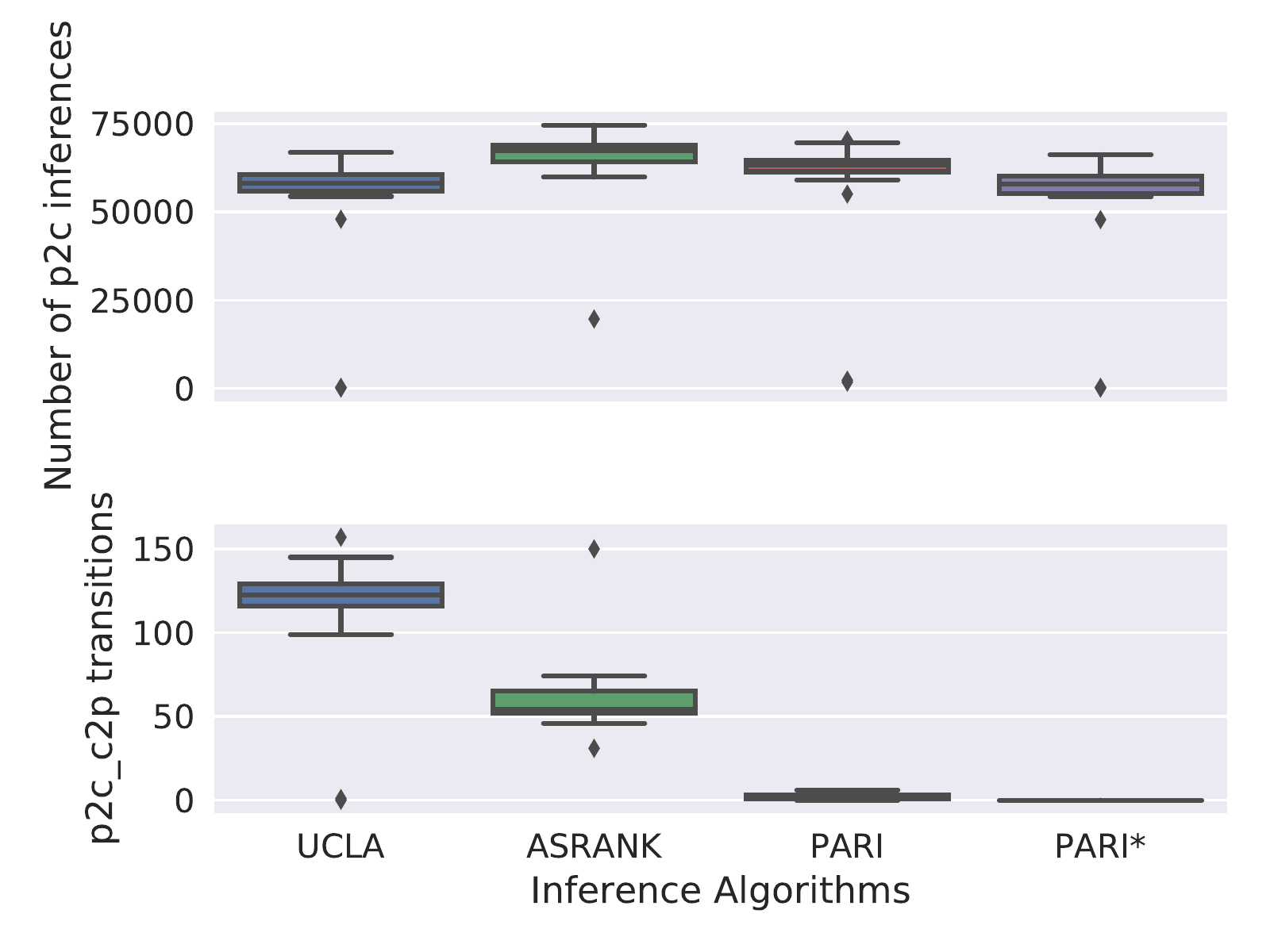}
	\caption{The upper plot shows the number of p2c (c2p) links inferred by \acrshort{ucla}, \acrshort{asrank}, \acrshort{pari}, and \acrshort{pari}* (a variant of \acrshort{pari}), when $|\overline{H}|=7$. PARI infers more p2c (c2p) links than \acrshort{ucla} but fewer than \acrshort{asrank}. The lower plot compares the number of transitions as $|\overline{H}|$ drops from 79 to 7 and illustrates the principle-based inference step's strong stability.}
	\label{fig:principle_p2c_compare_combined}
\end{figure}

The principle-based inference step not only attains higher accuracy than \acrshort{ucla} and \acrshort{asrank}, but also features greater stability when faced with dwindling visibility over the AS topology. The upper plot of Figure~\ref{fig:principle_p2c_compare_combined} compares the number of p2c inferences made by \acrshort{ucla}, \acrshort{asrank} and \acrshort{pari} when $|\overline{H}|$ (the set of active route collectors) is $7$. PARI infers more p2c (c2p) links than \acrshort{ucla} but fewer links than \acrshort{asrank}. As mentioned in~\S\ref{sec:principle-inf}, the principle-based inference step starts by inferring p2c links that follow the p2p links connecting two Tier-1 ASes. Once it gathers a set of p2c links, it repeatedly scans the AS paths to expand the link set based on inferences made in the prior iterations. It is this iterative process that allows PARI to infer more p2c links than UCLA, without compromising accuracy. We believe~\acrshort{asrank} infers more p2c links than~\acrshort{pari} because it incorporates a larger pool of rules, some of which are less reliable than the ones chosen by~\acrshort{pari}.

The lower plot of Figure~\ref{fig:principle_p2c_compare_combined} depicts the number of p2c\_c2p transitions as $|\overline{H}|$ drops from 79 to 7, following the procedure presented in Figure~\ref{fig:transitions}. We focus on p2c\_c2p transitions because the principle-based inference step only infers p2p relationship for links connecting Tier-1 ASes. Observe that \acrshort{pari} experiences many few transitions in comparison to \acrshort{ucla} and \acrshort{asrank}, exhibiting stronger stability. We also evaluate \acrshort{pari}*, a variant of \acrshort{pari} that omits the iterative p2c inference process. Surprisingly, the transitions can be further reduced to 0 at the expense of a minor drop in the number of p2c inferences. \acrshort{pari}'s advantage can be attributed to the decision to always categorize links with conflicting inferences as anomalies (i.e., links inferred as different types of relationship on multiple AS paths) instead of committing to one of them.

\subsection{Probabilistic Inference Results}
Experiments in this section only involve links that remain unclassified after \acrshort{pari}'s principle-based inference step. We identify the subset of these links that are labeled by the ground truth dataset and further divide this subset into training (70\%) and test (30\%) sets. For training, we use the \acrshort{mlr} implementation (LogisticRegressionCV) from scikit-learn~\cite{scikit-learn}, with cross-validation and L2 regularization to lessen the amount of overfitting. We report the performance of probabilistic inference by evaluating the probability estimates computed by the trained model over the test set. Unless stated otherwise, the training and test error rates, measured by cross-entropy loss, are within 5\% difference.

In order to compare the outputs of probabilistic inference and the deterministic algorithms, we convert every probability distribution into a single-relationship assignment. Specifically, we designate the type of relationship with the maximum probability as the single-relationship output of probabilistic inference. For example, consider the distribution $(0.06, 0.04, 0.9)^T$ where the elements are interpreted as the probability that the relationship is c2p, p2c, and p2p. The single-relationship output is p2p. It follows that conventional metrics such as accuracy becomes applicable. To evaluate an algorithm's performance, we measure the cross-entropy between the true and inferred probability distributions. The true distribution is obtained by applying 1-of-$K$ encoding~\cite{Bishop:2006:PRM:1162264} to the single-relationship assignment. For example, if a link is assigned p2p, we can associate it with the distribution $(0.0, 0.0, 1.0)^T$.

Below, we examine \acrshort{pari}'s probabilistic inference in terms of its accuracy of single-relationship inference, how well its final output probabilities reflect confidence and whether its share vectors are better than other heuristics for estimation of probabilities.

\subsubsection{Single-relationship Inference Accuracy}
For single-relationship inference, \acrshort{pari}'s probabilistic inference is comparable to the state-of-the-art algorithms. \acrshort{pari} achieves an accuracy of 93\% which is slightly lower than \acrshort{ucla} (94\%) and \acrshort{asrank} (96\%) but much higher than \acrshort{gao} (65\%). Note that we present the accuracy of single-relationship inference for completeness. The two key metrics to evaluate probabilistic inference are presented next.

\subsubsection{Probability Estimate Accuracy}
\begin{figure*}[h]
	\begin{subfigure}[b]{.49\textwidth}
		\centering\centering\captionsetup{width=.9\linewidth}
		\includegraphics[width=1.0\linewidth]{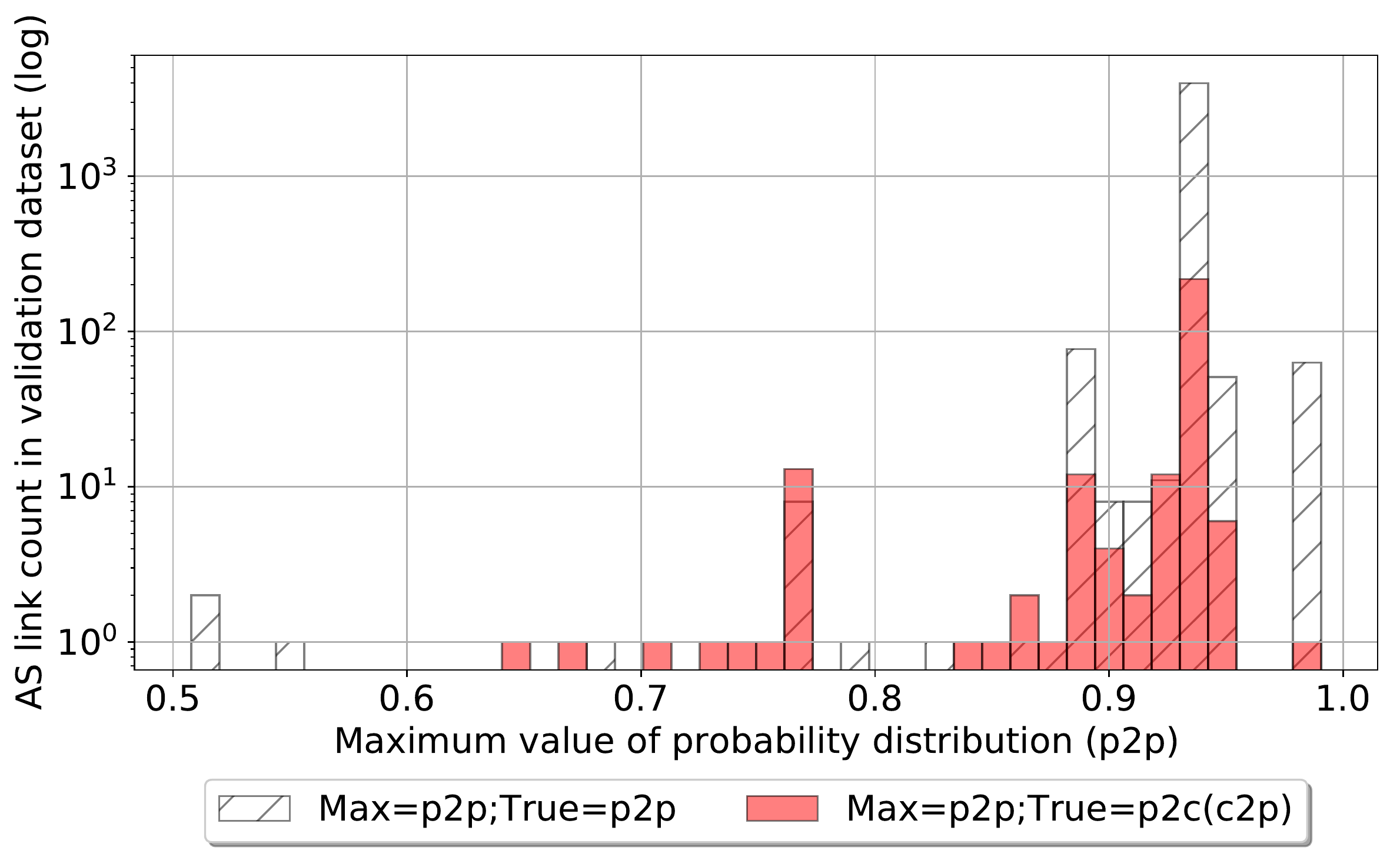}
		\subcaption{p2p}
		\label{fig:pari-max-true-p2p}
	\end{subfigure}%
	\begin{subfigure}[b]{.49\textwidth}
		\centering\centering\captionsetup{width=.9\linewidth}
		\includegraphics[width=\textwidth]{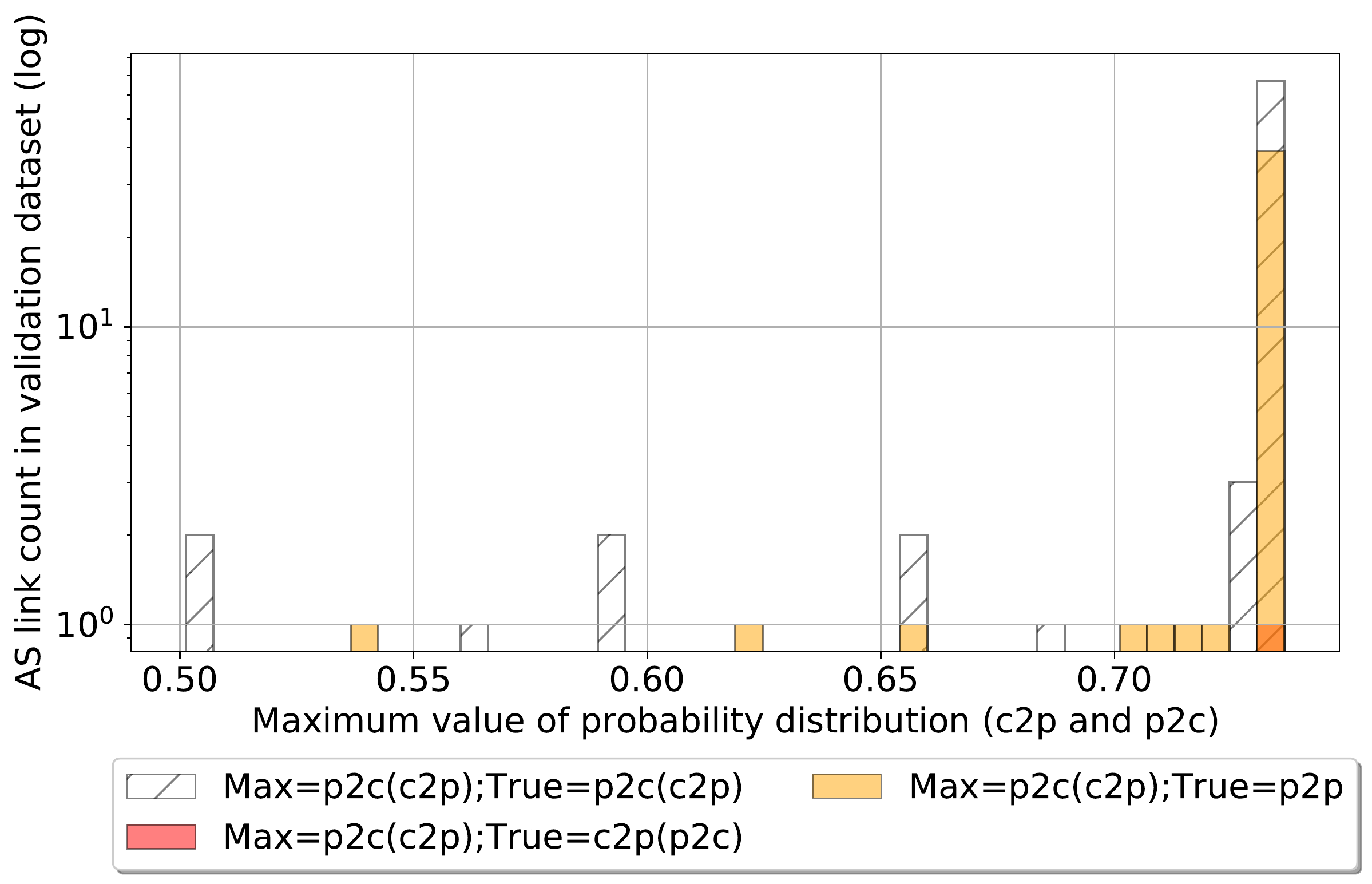}
		\subcaption{c2p/p2c}
		\label{fig:pari-max-true-no-p2p}
	\end{subfigure}
	\caption{Correlation between the relationship with the maximum probability estimate ($Max$) and the true relationship ($True$) for \acrshort{pari}. 
	}
\end{figure*}

One of \acrshort{pari}'s advantages over prior algorithms is its ability to effectively express a varying level of confidence in the inference output that is consistent with the ground-truth data. We illustrate it through the correlation between the relationship with the maximum probability estimate ($Max$) and the true relationship ($True$), of links in the test set. For links with $Max=p2p$ (shown in Figure~\ref{fig:pari-max-true-p2p}), we observe that the number of matches tend to dominate that of mismatches, indicating low uncertainty when the probability of p2p is high. In particular, in the rightmost bar (at 1.0), the number of matches at that point (hatched bar) is nearly two orders of magnitude higher than that of mismatches (red bar). In addition, For links with $Max=p2c$ or $c2p$, we recognize a similar trend in Figure~\ref{fig:pari-max-true-no-p2p} but the dominance of matches is less substantial.

\subsubsection{Efficacy of Share Vectors}
\begin{figure}[h]
	\centering
	\includegraphics[width=0.9\linewidth]{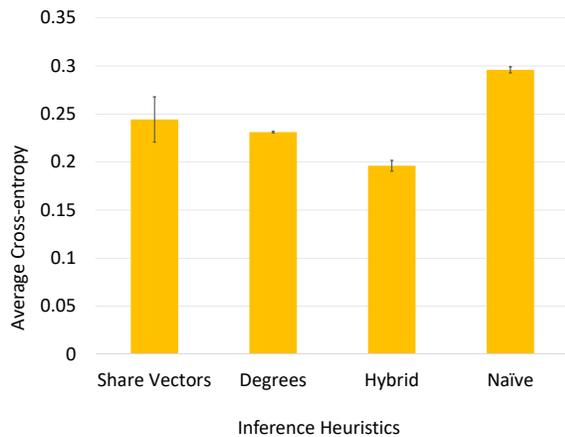}
	\caption{Average cross-entropy between the true and inferred probability distributions for four different heuristics and two deterministic algorithms. A Lower value implies smaller difference between the two distributions.}
	\label{fig:feature-compare}
\end{figure}

The share vectors are not the only heuristic that can be used to compute probability estimates; however, we believe that share vectors are effective compared to the alternatives. \acrshort{pari}'s key strength is that it opens the exploration of such algorithm designs. To illustrate the efficacy of share vectors, we consider three alternative for comparison: Na\"ive, Degrees and Hybrid. For any given link, the Na\"ive heuristic always outputs the relative frequency distribution of the true labels. For example, the probability of p2p is set to 0.92 if 92\% of the links from the input have p2p as the true label. The Degrees heuristic is a variant of \acrshort{pari}'s probabilistic inference step that replaces the link's share vector with the node/transit degrees of its endpoints as the link's feature. The last heuristic, Hybrid, includes both share vectors and node/transit degrees as features.

Figure~\ref{fig:feature-compare} measures the average cross-entropy between the true and inferred probability distributions for the above four heuristics, over 5 random splits of training and test sets. We examine the Na\"ive heuristic as a baseline for comparison. We note that all variants of \acrshort{pari} achieve lower average cross-entropy than Na\"ive. Degrees outperforms PARI with share vectors, showing that node/transit degrees is a useful feature.  Hybrid combines share vectors and node/transit degrees, two complementary features, and it outperforms both PARI with share vectors and Degrees. 
We believe Hybrid's advantage can be attributed to share vectors' ability to encapsulate the interdependence across neighbor links as imposed by the valley-free constraint, which cannot be easily captured by the local properties such as node/transit degrees.
	\section{Conclusion}
\label{sec:conclusion}
In this paper, we identify several causes of uncertainty in AS relationship inference and demonstrate why the problem deserves more attention by the research community. We propose a novel framework to inferring AS relationships that explicitly models the uncertainty and reflect it in inference output. We propose an exemplary algorithm (\acrshort{pari}) that implements this design paradigm and leverages a novel technique to capture the interdependence between undecided links. Our evaluation indicates it achieves higher accuracy and stability compared to prior algorithms for deterministic inference and outperforms a na\"ive strategy to express uncertainty.
	
	\bibliographystyle{ACM-Reference-Format}
	\bibliography{paper}

%%% -*-BibTeX-*-
%%% Do NOT edit. File created by BibTeX with style
%%% ACM-Reference-Format-Journals [18-Jan-2012].

\begin{thebibliography}{29}

%%% ====================================================================
%%% NOTE TO THE USER: you can override these defaults by providing
%%% customized versions of any of these macros before the \bibliography
%%% command.  Each of them MUST provide its own final punctuation,
%%% except for \shownote{}, \showDOI{}, and \showURL{}.  The latter two
%%% do not use final punctuation, in order to avoid confusing it with
%%% the Web address.
%%%
%%% To suppress output of a particular field, define its macro to expand
%%% to an empty string, or better, \unskip, like this:
%%%
%%% \newcommand{\showDOI}[1]{\unskip}   % LaTeX syntax
%%%
%%% \def \showDOI #1{\unskip}           % plain TeX syntax
%%%
%%% ====================================================================

\ifx \showCODEN    \undefined \def \showCODEN     #1{\unskip}     \fi
\ifx \showDOI      \undefined \def \showDOI       #1{#1}\fi
\ifx \showISBNx    \undefined \def \showISBNx     #1{\unskip}     \fi
\ifx \showISBNxiii \undefined \def \showISBNxiii  #1{\unskip}     \fi
\ifx \showISSN     \undefined \def \showISSN      #1{\unskip}     \fi
\ifx \showLCCN     \undefined \def \showLCCN      #1{\unskip}     \fi
\ifx \shownote     \undefined \def \shownote      #1{#1}          \fi
\ifx \showarticletitle \undefined \def \showarticletitle #1{#1}   \fi
\ifx \showURL      \undefined \def \showURL       {\relax}        \fi
% The following commands are used for tagged output and should be
% invisible to TeX
\providecommand\bibfield[2]{#2}
\providecommand\bibinfo[2]{#2}
\providecommand\natexlab[1]{#1}
\providecommand\showeprint[2][]{arXiv:#2}

\bibitem[\protect\citeauthoryear{Alaettinoglu, Villamizar, Gerich, Kessens,
  Meyer, Bates, Karrenberg, and Terpstra}{Alaettinoglu et~al\mbox{.}}{1999}]%
        {alaettinoglu1999routing}
\bibfield{author}{\bibinfo{person}{Cengiz Alaettinoglu},
  \bibinfo{person}{Curtis Villamizar}, \bibinfo{person}{Elise Gerich},
  \bibinfo{person}{David Kessens}, \bibinfo{person}{David Meyer},
  \bibinfo{person}{Tony Bates}, \bibinfo{person}{Daniel Karrenberg}, {and}
  \bibinfo{person}{Marten Terpstra}.} \bibinfo{year}{1999}\natexlab{}.
\newblock \bibinfo{booktitle}{\emph{Routing policy specification language
  (RPSL)}}.
\newblock \bibinfo{type}{{T}echnical {R}eport}.
\newblock


\bibitem[\protect\citeauthoryear{Biere, Heule, and van Maaren}{Biere
  et~al\mbox{.}}{2009}]%
        {biere2009handbook}
\bibfield{author}{\bibinfo{person}{Armin Biere}, \bibinfo{person}{Marijn
  Heule}, {and} \bibinfo{person}{Hans van Maaren}.}
  \bibinfo{year}{2009}\natexlab{}.
\newblock \bibinfo{booktitle}{\emph{Handbook of satisfiability}}.
  Vol.~\bibinfo{volume}{185}.
\newblock \bibinfo{publisher}{IOS press}.
\newblock


\bibitem[\protect\citeauthoryear{Bishop}{Bishop}{2006}]%
        {Bishop:2006:PRM:1162264}
\bibfield{author}{\bibinfo{person}{Christopher~M. Bishop}.}
  \bibinfo{year}{2006}\natexlab{}.
\newblock \bibinfo{booktitle}{\emph{Pattern Recognition and Machine Learning
  (Information Science and Statistics)}}.
\newblock \bibinfo{publisher}{Springer-Verlag}, \bibinfo{address}{Berlin,
  Heidelberg}.
\newblock
\showISBNx{0387310738}


\bibitem[\protect\citeauthoryear{Calder, Fan, Hu, Katz-Bassett, Heidemann, and
  Govindan}{Calder et~al\mbox{.}}{2013}]%
        {calder2013mapping}
\bibfield{author}{\bibinfo{person}{Matt Calder}, \bibinfo{person}{Xun Fan},
  \bibinfo{person}{Zi Hu}, \bibinfo{person}{Ethan Katz-Bassett},
  \bibinfo{person}{John Heidemann}, {and} \bibinfo{person}{Ramesh Govindan}.}
  \bibinfo{year}{2013}\natexlab{}.
\newblock \showarticletitle{Mapping the expansion of Google's serving
  infrastructure}. In \bibinfo{booktitle}{\emph{Proceedings of the 2013
  conference on Internet measurement conference}}. ACM,
  \bibinfo{pages}{313--326}.
\newblock


\bibitem[\protect\citeauthoryear{Center}{Center}{2017}]%
        {ripe}
\bibfield{author}{\bibinfo{person}{RIPE Network~Coordination Center}.}
  \bibinfo{year}{2017}\natexlab{}.
\newblock \bibinfo{title}{Routing Information Service Raw Data}.
\newblock
  \bibinfo{howpublished}{\url{https://www.ripe.net/analyse/internet-measurements/routing-information-service-ris/ris-raw-data}}.
\newblock
\newblock
\shownote{[Online; accessed 12-October-2017].}


\bibitem[\protect\citeauthoryear{Chandra, Traina, and Li}{Chandra
  et~al\mbox{.}}{1996}]%
        {chandra1996bgp}
\bibfield{author}{\bibinfo{person}{R Chandra}, \bibinfo{person}{P Traina},
  {and} \bibinfo{person}{T Li}.} \bibinfo{year}{1996}\natexlab{}.
\newblock \bibinfo{booktitle}{\emph{BGP communities attribute}}.
\newblock \bibinfo{type}{{T}echnical {R}eport}.
\newblock


\bibitem[\protect\citeauthoryear{Di~Battista, Patrignani, and
  Pizzonia}{Di~Battista et~al\mbox{.}}{2003}]%
        {di2003computing}
\bibfield{author}{\bibinfo{person}{Giuseppe Di~Battista},
  \bibinfo{person}{Maurizio Patrignani}, {and} \bibinfo{person}{Maurizio
  Pizzonia}.} \bibinfo{year}{2003}\natexlab{}.
\newblock \showarticletitle{Computing the types of the relationships between
  autonomous systems}. In \bibinfo{booktitle}{\emph{INFOCOM 2003. Twenty-Second
  Annual Joint Conference of the IEEE Computer and Communications. IEEE
  Societies}}, Vol.~\bibinfo{volume}{1}. IEEE, \bibinfo{pages}{156--165}.
\newblock


\bibitem[\protect\citeauthoryear{Dimitropoulos, Krioukov, Fomenkov, Huffaker,
  Hyun, Riley, et~al\mbox{.}}{Dimitropoulos et~al\mbox{.}}{2007}]%
        {dimitropoulos2007relationships}
\bibfield{author}{\bibinfo{person}{Xenofontas Dimitropoulos},
  \bibinfo{person}{Dmitri Krioukov}, \bibinfo{person}{Marina Fomenkov},
  \bibinfo{person}{Bradley Huffaker}, \bibinfo{person}{Young Hyun},
  \bibinfo{person}{George Riley}, {et~al\mbox{.}}}
  \bibinfo{year}{2007}\natexlab{}.
\newblock \showarticletitle{AS relationships: Inference and validation}.
\newblock \bibinfo{journal}{\emph{ACM SIGCOMM Computer Communication Review}}
  \bibinfo{volume}{37}, \bibinfo{number}{1} (\bibinfo{year}{2007}),
  \bibinfo{pages}{29--40}.
\newblock


\bibitem[\protect\citeauthoryear{Gao}{Gao}{2001}]%
        {gao2001inferring}
\bibfield{author}{\bibinfo{person}{Lixin Gao}.}
  \bibinfo{year}{2001}\natexlab{}.
\newblock \showarticletitle{On inferring autonomous system relationships in the
  Internet}.
\newblock \bibinfo{journal}{\emph{IEEE/ACM Transactions on Networking (ToN)}}
  \bibinfo{volume}{9}, \bibinfo{number}{6} (\bibinfo{year}{2001}),
  \bibinfo{pages}{733--745}.
\newblock


\bibitem[\protect\citeauthoryear{Garey and Johnson}{Garey and Johnson}{1979}]%
        {garey1979computers}
\bibfield{author}{\bibinfo{person}{Michael~R Garey} {and}
  \bibinfo{person}{David~S Johnson}.} \bibinfo{year}{1979}\natexlab{}.
\newblock \bibinfo{title}{Computers and intractability. A guide to the theory
  of NP-completeness. A Series of Books in the Mathematical Sciences}.
\newblock
\newblock


\bibitem[\protect\citeauthoryear{Giotsas and Zhou}{Giotsas and Zhou}{2012}]%
        {giotsas2012valley}
\bibfield{author}{\bibinfo{person}{Vasileios Giotsas} {and}
  \bibinfo{person}{Shi Zhou}.} \bibinfo{year}{2012}\natexlab{}.
\newblock \showarticletitle{Valley-free violation in Internet
  routing—Analysis based on BGP Community data}. In
  \bibinfo{booktitle}{\emph{Communications (ICC), 2012 IEEE International
  Conference on}}. IEEE, \bibinfo{pages}{1193--1197}.
\newblock


\bibitem[\protect\citeauthoryear{House}{House}{2017}]%
        {pch}
\bibfield{author}{\bibinfo{person}{Packet~Clearing House}.}
  \bibinfo{year}{2017}\natexlab{}.
\newblock \bibinfo{title}{Packet Clearing House Daily Routing Snapshots}.
\newblock
  \bibinfo{howpublished}{\url{https://www.pch.net/resources/Routing_Data/}}.
\newblock
\newblock
\shownote{[Online; accessed 12-October-2017].}


\bibitem[\protect\citeauthoryear{Huang, Li, and Ross}{Huang
  et~al\mbox{.}}{2007}]%
        {huang2007can}
\bibfield{author}{\bibinfo{person}{Cheng Huang}, \bibinfo{person}{Jin Li},
  {and} \bibinfo{person}{Keith~W Ross}.} \bibinfo{year}{2007}\natexlab{}.
\newblock \showarticletitle{Can internet video-on-demand be profitable?}
\newblock \bibinfo{journal}{\emph{ACM SIGCOMM Computer Communication Review}}
  \bibinfo{volume}{37}, \bibinfo{number}{4} (\bibinfo{year}{2007}),
  \bibinfo{pages}{133--144}.
\newblock


\bibitem[\protect\citeauthoryear{Luckie, Dhamdhere, Clark, Huffaker,
  et~al\mbox{.}}{Luckie et~al\mbox{.}}{2014}]%
        {luckie2014challenges}
\bibfield{author}{\bibinfo{person}{Matthew Luckie}, \bibinfo{person}{Amogh
  Dhamdhere}, \bibinfo{person}{David Clark}, \bibinfo{person}{Bradley
  Huffaker}, {et~al\mbox{.}}} \bibinfo{year}{2014}\natexlab{}.
\newblock \showarticletitle{Challenges in inferring internet interdomain
  congestion}. In \bibinfo{booktitle}{\emph{Proceedings of the 2014 Conference
  on Internet Measurement Conference}}. ACM, \bibinfo{pages}{15--22}.
\newblock


\bibitem[\protect\citeauthoryear{Luckie, Huffaker, Dhamdhere, Giotsas,
  et~al\mbox{.}}{Luckie et~al\mbox{.}}{2013}]%
        {luckie2013relationships}
\bibfield{author}{\bibinfo{person}{Matthew Luckie}, \bibinfo{person}{Bradley
  Huffaker}, \bibinfo{person}{Amogh Dhamdhere}, \bibinfo{person}{Vasileios
  Giotsas}, {et~al\mbox{.}}} \bibinfo{year}{2013}\natexlab{}.
\newblock \showarticletitle{AS relationships, customer cones, and validation}.
  In \bibinfo{booktitle}{\emph{Proceedings of the 2013 conference on Internet
  measurement conference}}. ACM, \bibinfo{pages}{243--256}.
\newblock


\bibitem[\protect\citeauthoryear{Mahajan, Wetherall, and Anderson}{Mahajan
  et~al\mbox{.}}{2002}]%
        {mahajan2002understanding}
\bibfield{author}{\bibinfo{person}{Ratul Mahajan}, \bibinfo{person}{David
  Wetherall}, {and} \bibinfo{person}{Tom Anderson}.}
  \bibinfo{year}{2002}\natexlab{}.
\newblock \showarticletitle{Understanding BGP misconfiguration}. In
  \bibinfo{booktitle}{\emph{ACM SIGCOMM Computer Communication Review}},
  Vol.~\bibinfo{volume}{32}. ACM, \bibinfo{pages}{3--16}.
\newblock


\bibitem[\protect\citeauthoryear{Martins, Manquinho, and Lynce}{Martins
  et~al\mbox{.}}{2014}]%
        {martins2014open}
\bibfield{author}{\bibinfo{person}{Ruben Martins}, \bibinfo{person}{Vasco
  Manquinho}, {and} \bibinfo{person}{In{\^e}s Lynce}.}
  \bibinfo{year}{2014}\natexlab{}.
\newblock \showarticletitle{Open-WBO: A modular MaxSAT solver}. In
  \bibinfo{booktitle}{\emph{International Conference on Theory and Applications
  of Satisfiability Testing}}. Springer, \bibinfo{pages}{438--445}.
\newblock


\bibitem[\protect\citeauthoryear{Niu, R{\'e}, Doan, and Shavlik}{Niu
  et~al\mbox{.}}{2011}]%
        {niu2011tuffy}
\bibfield{author}{\bibinfo{person}{Feng Niu}, \bibinfo{person}{Christopher
  R{\'e}}, \bibinfo{person}{AnHai Doan}, {and} \bibinfo{person}{Jude Shavlik}.}
  \bibinfo{year}{2011}\natexlab{}.
\newblock \showarticletitle{Tuffy: Scaling up statistical inference in markov
  logic networks using an rdbms}.
\newblock \bibinfo{journal}{\emph{Proceedings of the VLDB Endowment}}
  \bibinfo{volume}{4}, \bibinfo{number}{6} (\bibinfo{year}{2011}),
  \bibinfo{pages}{373--384}.
\newblock


\bibitem[\protect\citeauthoryear{of~Oregon}{of~Oregon}{2017}]%
        {routeviews}
\bibfield{author}{\bibinfo{person}{University of Oregon}.}
  \bibinfo{year}{2017}\natexlab{}.
\newblock \bibinfo{title}{University of Oregon Route Views Project}.
\newblock \bibinfo{howpublished}{\url{http://www.routeviews.org/}}.
\newblock
\newblock
\shownote{[Online; accessed 12-October-2017].}


\bibitem[\protect\citeauthoryear{Oliveira, Pei, Willinger, Zhang, and
  Zhang}{Oliveira et~al\mbox{.}}{2010}]%
        {oliveira2010completeness}
\bibfield{author}{\bibinfo{person}{Ricardo Oliveira}, \bibinfo{person}{Dan
  Pei}, \bibinfo{person}{Walter Willinger}, \bibinfo{person}{Beichuan Zhang},
  {and} \bibinfo{person}{Lixia Zhang}.} \bibinfo{year}{2010}\natexlab{}.
\newblock \showarticletitle{The (in) completeness of the observed internet
  AS-level structure}.
\newblock \bibinfo{journal}{\emph{IEEE/ACM Transactions on Networking (ToN)}}
  \bibinfo{volume}{18}, \bibinfo{number}{1} (\bibinfo{year}{2010}),
  \bibinfo{pages}{109--122}.
\newblock


\bibitem[\protect\citeauthoryear{Oliveira, Pei, Willinger, Zhang, and
  Zhang}{Oliveira et~al\mbox{.}}{2008}]%
        {oliveira2008search}
\bibfield{author}{\bibinfo{person}{Ricardo~V Oliveira}, \bibinfo{person}{Dan
  Pei}, \bibinfo{person}{Walter Willinger}, \bibinfo{person}{Beichuan Zhang},
  {and} \bibinfo{person}{Lixia Zhang}.} \bibinfo{year}{2008}\natexlab{}.
\newblock \showarticletitle{In search of the elusive ground truth: the
  internet's as-level connectivity structure}. In \bibinfo{booktitle}{\emph{ACM
  SIGMETRICS Performance Evaluation Review}}, Vol.~\bibinfo{volume}{36}. ACM,
  \bibinfo{pages}{217--228}.
\newblock


\bibitem[\protect\citeauthoryear{Pedregosa, Varoquaux, Gramfort, Michel,
  Thirion, Grisel, Blondel, Prettenhofer, Weiss, Dubourg, Vanderplas, Passos,
  Cournapeau, Brucher, Perrot, and Duchesnay}{Pedregosa et~al\mbox{.}}{2011}]%
        {scikit-learn}
\bibfield{author}{\bibinfo{person}{F. Pedregosa}, \bibinfo{person}{G.
  Varoquaux}, \bibinfo{person}{A. Gramfort}, \bibinfo{person}{V. Michel},
  \bibinfo{person}{B. Thirion}, \bibinfo{person}{O. Grisel},
  \bibinfo{person}{M. Blondel}, \bibinfo{person}{P. Prettenhofer},
  \bibinfo{person}{R. Weiss}, \bibinfo{person}{V. Dubourg}, \bibinfo{person}{J.
  Vanderplas}, \bibinfo{person}{A. Passos}, \bibinfo{person}{D. Cournapeau},
  \bibinfo{person}{M. Brucher}, \bibinfo{person}{M. Perrot}, {and}
  \bibinfo{person}{E. Duchesnay}.} \bibinfo{year}{2011}\natexlab{}.
\newblock \showarticletitle{{Scikit-learn: Machine Learning in Python }}.
\newblock \bibinfo{journal}{\emph{Journal of Machine Learning Research}}
  \bibinfo{volume}{12} (\bibinfo{year}{2011}), \bibinfo{pages}{2825--2830}.
\newblock


\bibitem[\protect\citeauthoryear{Richardson and Domingos}{Richardson and
  Domingos}{2006}]%
        {richardson2006markov}
\bibfield{author}{\bibinfo{person}{Matthew Richardson} {and}
  \bibinfo{person}{Pedro Domingos}.} \bibinfo{year}{2006}\natexlab{}.
\newblock \showarticletitle{Markov logic networks}.
\newblock \bibinfo{journal}{\emph{Machine learning}} \bibinfo{volume}{62},
  \bibinfo{number}{1-2} (\bibinfo{year}{2006}), \bibinfo{pages}{107--136}.
\newblock


\bibitem[\protect\citeauthoryear{Roughan, Willinger, Maennel, Perouli, and
  Bush}{Roughan et~al\mbox{.}}{2011}]%
        {roughan201110}
\bibfield{author}{\bibinfo{person}{Matthew Roughan}, \bibinfo{person}{Walter
  Willinger}, \bibinfo{person}{Olaf Maennel}, \bibinfo{person}{Debbie Perouli},
  {and} \bibinfo{person}{Randy Bush}.} \bibinfo{year}{2011}\natexlab{}.
\newblock \showarticletitle{10 lessons from 10 years of measuring and modeling
  the internet's autonomous systems}.
\newblock \bibinfo{journal}{\emph{IEEE Journal on Selected Areas in
  Communications}} \bibinfo{volume}{29}, \bibinfo{number}{9}
  (\bibinfo{year}{2011}), \bibinfo{pages}{1810--1821}.
\newblock


\bibitem[\protect\citeauthoryear{Shah and Zaman}{Shah and Zaman}{2011}]%
        {shah2011rumors}
\bibfield{author}{\bibinfo{person}{Devavrat Shah} {and} \bibinfo{person}{Tauhid
  Zaman}.} \bibinfo{year}{2011}\natexlab{}.
\newblock \showarticletitle{Rumors in a network: Who's the culprit?}
\newblock \bibinfo{journal}{\emph{IEEE Transactions on information theory}}
  \bibinfo{volume}{57}, \bibinfo{number}{8} (\bibinfo{year}{2011}),
  \bibinfo{pages}{5163--5181}.
\newblock


\bibitem[\protect\citeauthoryear{Subramanian, Agarwal, Rexford, and
  Katz}{Subramanian et~al\mbox{.}}{2002}]%
        {subramanian2002characterizing}
\bibfield{author}{\bibinfo{person}{Lakshminarayanan Subramanian},
  \bibinfo{person}{Sharad Agarwal}, \bibinfo{person}{Jennifer Rexford}, {and}
  \bibinfo{person}{Randy~H Katz}.} \bibinfo{year}{2002}\natexlab{}.
\newblock \showarticletitle{Characterizing the Internet hierarchy from multiple
  vantage points}. In \bibinfo{booktitle}{\emph{INFOCOM 2002. Twenty-First
  Annual Joint Conference of the IEEE Computer and Communications Societies.
  Proceedings. IEEE}}, Vol.~\bibinfo{volume}{2}. IEEE,
  \bibinfo{pages}{618--627}.
\newblock


\bibitem[\protect\citeauthoryear{THURLEY}{THURLEY}{2006}]%
        {thurley2006sharpsat}
\bibfield{author}{\bibinfo{person}{Marc THURLEY}.}
  \bibinfo{year}{2006}\natexlab{}.
\newblock \showarticletitle{sharpSAT: Counting models with advanced component
  caching and implicit BCP}.
\newblock \bibinfo{journal}{\emph{Lecture notes in computer science}}
  (\bibinfo{year}{2006}), \bibinfo{pages}{424--429}.
\newblock


\bibitem[\protect\citeauthoryear{tier-one}{tier-one}{2017}]%
        {tier-one}
tier-one \bibinfo{year}{2017}\natexlab{}.
\newblock \bibinfo{title}{Tier 1 network}.
\newblock
  \bibinfo{howpublished}{\url{https://en.wikipedia.org/wiki/Tier_1_network}}.
\newblock
\newblock
\shownote{[Online; accessed 12-October-2017].}


\bibitem[\protect\citeauthoryear{Xia and Gao}{Xia and Gao}{2004}]%
        {xia2004evaluation}
\bibfield{author}{\bibinfo{person}{Jianhong Xia} {and} \bibinfo{person}{Lixin
  Gao}.} \bibinfo{year}{2004}\natexlab{}.
\newblock \showarticletitle{On the evaluation of AS relationship inferences
  [Internet reachability/traffic flow applications]}. In
  \bibinfo{booktitle}{\emph{Global Telecommunications Conference, 2004.
  GLOBECOM'04. IEEE}}, Vol.~\bibinfo{volume}{3}. IEEE,
  \bibinfo{pages}{1373--1377}.
\newblock


\end{thebibliography}
	
\end{document}